%% file: main.tex
\documentclass[12pt,a4paper]{article}
\pdfoutput=1
\usepackage{graphicx,epsfig}
\usepackage{dcolumn} 
\usepackage{slashed,color,amsmath,amssymb}
\usepackage{jheppub}
\usepackage{enumitem}
\usepackage{siunitx}
\usepackage{bm}
\usepackage{layouts}
\usepackage[svgnames]{xcolor}
\usepackage[normalem]{ulem}
\usepackage{subfiles}
\usepackage{caption, subcaption}
\usepackage{multirow}
\usepackage{verbatim}
\usepackage[normalem]{ulem}

\usepackage{soul}
\newcommand{\be}{\begin{equation}}
\newcommand{\ee}{\end{equation}}
\newcommand{\bea}{\begin{eqnarray}}
\newcommand{\eea}{\end{eqnarray}}
\newcommand{\bit}{\begin{itemize}}
\newcommand{\eit}{\end{itemize}}

\newcommand{\lam}{\lambda}

\definecolor{newgreen}{HTML}{009900}

\newcommand\thefontsize{The current font size is: \f@size pt}

  \definecolor{fsred}{HTML}{c71616}

\definecolor{fsblue}{HTML}{1e88e5}

\definecolor{fsyellow}{HTML}{dca607}

\definecolor{fsgreen}{HTML}{014d40}

\definecolor{fsviolet}{HTML}{a11b9f}

\definecolor{newpurple}{HTML}{7B1DAA}

\def\gsim{\lower0.5ex\hbox{$\:\buildrel >\over\sim\:$}}
\def\lsim{\lower0.5ex\hbox{$\:\buildrel <\over\sim\:$}}

\hypersetup{
   colorlinks=true,       
   linkcolor=blue,        
   citecolor=red,         
   filecolor=magenta,     
}

\preprint{\begin{flushright} BONN-TH-2025-22, IPPP/25/48 
	\end{flushright}}	  

\title{Colliders are not Testing Locality via Bell's Inequality nor Providing an 
Unconditional Proof of Entanglement}

        \author[a]{Steven A. Abel,}
	\emailAdd{steve.abel@durham.ac.uk}
	\affiliation[a]{Institute for Particle Physics Phenomenology, and Department of Mathematical Sciences, Durham University, Durham DH1 3LE, UK}

        \author[b]{Herbi~K.~Dreiner,}
	\emailAdd{dreiner@uni-bonn.de}
	\affiliation[b]{Bethe Center for Theoretical Physics \& Physikalisches Institut der Universit\"at Bonn,\\ Nu{\ss}allee 12, 53115 Bonn, Germany}
					
	\author[b]{Rhitaja Sengupta,}
	\emailAdd{rsengupt@uni-bonn.de}
		
	\author[c]{Lorenzo Ubaldi}
	\emailAdd{lorenzo.ubaldi@ijs.si}
        \affiliation[c]{Jožef Stefan Institute, Jamova 39, 1000 Ljubljana, Slovenia}

\abstract{
Recently there has been an increased interest in possible tests of locality via 
Bell's inequalities, as well as separately tests of entanglement at colliders, 
in particular at the LHC. These have involved various physical processes, such 
as $t\bar t$, or $\tau^+\tau^-$ production, or the decay of a Higgs boson to 
two vector bosons $H\to VV^*$. We argue that \textit{none} of these proposals 
constitute a test of locality via Bell's inequality or promise 
unconditional observational evidence of entanglement. In all cases what 
is measured are the momenta of the final state particles. Using the construction 
proposed by Kasday (1971) in a different context, and adapted to collider 
scenarios by Abel, Dittmar, and Dreiner (1992), it is 
straightforward to construct a local hidden variable theory (LHVT) which 
exactly reproduces the data. This construction is only possible as the final 
state momenta all commute. We show that this LHVT satisfies Bell's 
inequality or the related CHSH inequality as appropriate for all the proposed 
LHC collider tests in the literature. Thus a test of locality via Bell's 
inequality is not possible at colliders. The LHVT is also by construction 
local, \textit{i.e.} all correlations are separable. Thus an unconditional proof 
of entanglement is also inherently \textit{not} 
possible at colliders. It can only be shown that the entanglement within the 
Standard Model consistently describes the data.
}

\begin{document}

\maketitle 


\subfile{tex/01_introduction/introduction}


\subfile{tex/02_tau-tau-LEP/Our-1992-Paper}

\subfile{tex/04_analysis/analysis}



\subfile{tex/05_cuts_and_Bell/cuts_and_Bell}


\section*{Acknowledgement}
HKD would like to thank the Universiteit van Amsterdam for a visiting professorship,
as well as my host during my stay, the Nikhef theory group. Part of this work was 
presented there as a seminar, which lead to very lively discussions, which are 
hopefully properly taken into account here. We would also like to thank Alan Barr, 
Philip Bechtle, Katharina Behr, Cedric Breuning, Klaus Desch, Jordi de Vries, Michael 
Dittmar, Lance Dixon, Manuel Drees, Claude Duhr, John Ellis, Marumi Kado, Chris 
Korthals Altes, Eric Laenen, Johannes Michel, Marieke Postma, Simon Sanders, Xerxes 
Tata, Georg Weiglein, and Dieter Zeppenfeld. HKD would like to thank theory group at 
the Jožef Stefan Institute, Ljubljana, in particular Jernej Fesel Kamenik, as well as 
the participants at the UK Annual Theory Meeting in Durham for enriching discussions, 
in particular the organizer Michael Spannowsky. SAA is supported by the STFC under 
Grant No. ST/T001011/1. LU is supported by the Slovenian Research Agency under the 
research core funding No.P1-0035, and by the research grants J1-60026 and J1-4389.


\appendix

\subfile{tex/A1_appendix_separability.tex}

\subfile{tex/04_analysis/analysis_Higgs_new}

\bibliographystyle{JHEP}

\providecommand{\href}[2]{#2}\begingroup\raggedright\endgroup

\end{document}

%% file: tex/01_introduction/introduction.tex

\section{Introduction}
In his 1964 paper Bell derived an inequality which enabled the experimental test 
of the question of locality versus quantum mechanics \cite{Bell:1964kc}. The 
inequality was expressed based on a Gedankenexperiment for the correlation 
functions of a pair of correlated or entangled spin-1/2 states first proposed
by Bohm and Aharonov \cite{Bohm:1957zz}, as a variation of the Gedankenexperiment
of Einstein, Podolsky, and Rosen \cite{Einstein:1935rr}. Clauser, Horne, Shimony, and Holt 
generalized the inequality to apply it to realizable experiments 
\cite{Clauser:1969ny}, the so-called CHSH inequality. In particular they picked up
an earlier proposal also by Bohm and Aharonov \cite{Bohm:1957zz} to perform tests on 
correlated spin-1 photon pairs instead of spin-1/2 fermions. This test was finally 
successfully performed by Aspect and collaborators with a resounding success for 
quantum mechanics \cite{Aspect:1982fx}.

For many people this has settled the question. However, the Aspect experiments were 
not performed on fermion pairs as in the Bell Gedankenexperiment and were 
restricted to photons in the eV energy range. Could it be that locality and 
entanglement behave differently at higher energies or correspondingly shorter length
scales? And could fermions be different from bosons in this context? Recently there 
has been a resurgent interest in potential experimental tests of locality via 
Bell's inequality and/or of entanglement at colliders, in particular at the LHC, and 
at SuperKEKB, \textit{i.e.,} at very high energies compared to the Aspect experiments. 
Most notable have been the extensive proposals related to $t\bar t$ production at 
the LHC and future colliders \textit{e.g.,} Refs.~\cite{Afik:2020onf, Fabbrichesi:2021npl,  
Severi:2021cnj, 
Aoude:2022imd,
Afik:2022kwm, 
Aguilar-Saavedra:2022uye, 
Fabbrichesi:2022ovb,
Afik:2022dgh,
Severi:2022qjy,
Dong:2023xiw,
Aguilar-Saavedra:2023hss,
Han:2023fci,
Cheng:2023qmz, 
Aguilar-Saavedra:2024fig,
Maltoni:2024tul,
Aguilar-Saavedra:2024hwd,
Aguilar-Saavedra:2024vpd,
Maltoni:2024csn,
White:2024nuc,
Dong:2024xsg,
Cheng:2024btk,
Dong:2024xsb,
Han:2024ugl}, 
as well as the measurements by ATLAS and CMS claiming to only test for entanglement 
at the hitherto highest possible energies \cite{ATLAS:2023fsd, CMS:2024pts, CMS:2024zkc}. 
Here the top-spin correlations should be determined via the correlated momenta of the leptons 
in the leptonic decay of each top, respectively. Note that CMS themselves has recently put 
forward an opposite opinion regarding the logical viability of such tests \cite{CMS:2025anw}.
In a similar context, $\tau^+\tau^-$ production \cite{Altakach:2022ywa, Ehataht:2023zzt, 
LoChiatto:2024dmx, Breuning:2024wxg, Han:2025ewp, Zhang:2025mmm} and $b\bar b$ production 
\cite{Afik:2025grr, Kats:2023zxb} at 
the LHC have been studied. There have also been related proposals to test for a violation of 
Bell locality and/or for entanglement in the spin correlations of spin-1 gauge bosons for 
example in decays of the Higgs boson $H^0\to W^+ W^-, Z^0Z^0$ 
\cite{Barr:2021zcp, Barr:2022wyq, Aguilar-Saavedra:2022wam, Ashby-Pickering:2022umy, 
Aguilar-Saavedra:2022mpg, Fabbrichesi:2023cev, 
Aoude:2023hxv,
Bernal:2023ruk,
Bi:2023uop,
Aoude:2023hxv, 
Aguilar-Saavedra:2024whi,
Subba:2024mnl,
Ruzi:2024cbt,
Grossi:2024jae,
Sullivan:2024wzl,
Wu:2024ovc,
DelGratta:2025qyp}, see also \cite{Grabarczyk:2024wnk, ATLAS:2026hye}.

In an earlier paper two of us together with Dittmar analyzed $\tau^+\tau^-$ production on the 
$Z^0$ resonance at LEP, followed by the decays $\tau^\pm\to\pi^\pm\nu$ \cite{Abel:1992kz, 
Dreiner:1992gt}. On fairly general grounds we showed that it is \textit{not} possible to test 
locality via Bell's inequality in such processes. The basic argument was that, since we measure 
final state 4-momenta at colliders and since momenta all commute, the differential cross section, 
in this case as a function of the final state pion momenta, itself constitutes a local hidden 
variable theory (LHVT) in the sense of Bell \cite{Bell:1964kc}. This is an application of the 
work by Kasday~\cite{kasday1971experimental} to the collider context. LHVTs satisfy Bell's 
inequality (or correspondingly the CHSH inequality) and are inherently local and thus separable\,\footnote{For clarity, we discuss in detail the definition of separability for classical LHVTS in Appendix~\ref{app:separability}.}. 
This could be considered a no-go theorem~\cite{Bechtle:2025ugc}.

The analysis and conclusions of Ref.~\cite{Abel:1992kz} have subsequently been largely
overlooked.\!\footnote{Note Ref.~\cite{Li:2024luk} which is at least partially in the spirit 
of our paper. However in contrast, we find that no meaningful tests of LHVTs can be performed. 
See also Ref.~\cite{Breuning:2024wxg}.} In particular its line of thought was not taken into 
account in the case of the more recent experimental proposals and measurements. It is the 
purpose of this paper to remedy this situation and to extend our previous analysis to the 
LHC scenarios. We show that in all the above mentioned cases the relevant differential cross 
section is again an LHVT. For the di-fermion production we show that it satisfies Bell's 
inequality or if appropriate the CHSH inequality, and the data can thus be described by an 
inherently local and separable function ({\it cf.}~Appendix~\ref{app:separability}). There can thus be no test of locality via Bell's 
inequality nor separately an observational method for colliders to eliminate the possibility 
of an entirely separable classical explanation of the results. There can thus be no test of 
locality via Bell's inequality, and any observation consistent with entangled top-quark pairs 
in the Standard Model remains compatible with local hidden-variable theories, precluding a 
conclusive test of entanglement. For a further recent related critical discussion see 
Ref.~\cite{Bechtle:2025ugc}. There it is shown that further theory input, \textit{e.g.} 
the Standard Model, is required to extract the underlying spin correlation, for example of the 
top quarks, from the observed angular correlations of momenta, in the $t\bar t$ case of the 
charged leptons.

The outline of this paper is as follows. In Section~2, we review the argument given in our 
paper from 1992, which is an application of a construction by Kasday 
\cite{kasday1971experimental} to the collider setup, here the LEP collider. In Section~3, we 
extend this argument to some other considered collider reactions involving fermions, including the $t\bar t$-production scenario at the LHC. In Section~4, we discuss 
the case of Higgs boson decays to massive vector bosons.
In Section~5, we show how phase space cuts 
(filters) can mistakenly lead to correlation functions which violate Bell's 
inequality. Finally in Section~6, we summarize and conclude. 
In Appendix~\ref{app:separability}, we discuss the definition of separability for both quantum systems and the classical LHVTs, that is used throughout the paper.
Lastly, in Appendix~\ref{app:massive_spin1_Bell}, 
we discuss the special case of the Higgs decay to transversely polarized massive vector bosons, $H\to V_TV^*_T$.

%% file: tex/02_tau-tau-LEP/Our-1992-Paper.tex
\section{Possible Test at LEP: $e^+e^-\to Z^0\to \tau^+\tau^-\to(\pi^+\bar\nu_\tau)
(\pi^-\nu_\tau)$}
\label{sec:previous_work}

Here we briefly review our paper from 1992 \cite{Abel:1992kz}, as the line of 
argument we employ in our critique of the more recent proposals at the LHC is 
similar. That work considered a possible test of locality via Bell's inequality at 
the LEP collider via the process
\begin{equation}
    e^+e^-\to Z^0 \to \tau^+\tau^- \to (\pi^+\bar\nu_\tau)(\pi^-\nu_\tau)\,.
\end{equation}
In this reaction the $\tau$ spins from the $Z^0$ decay are correlated. The $\tau$'s 
decay via the weak interactions in a so-called self-analyzing decay, and thus the 
pion momenta are also correlated. It is useful to consider the pion momenta in 
their respective parent $\tau$ rest frame. As it is a 2-body decay, the energies of 
the pions in these respective rest-frames are fixed, and we need only consider the 
unit pion momenta, which we denote by $\hat{\mathbf{p}}_{\pi^\pm}$, respectively. The 
differential cross section at leading order in these special frames is computed to 
be
\begin{equation}
    \frac{d\sigma}{d\cos\theta_{\pi\pi}}(e^+e^-\to\pi^+\bar\nu_\tau\pi^-\nu_\tau) =
    A (1-\frac{1}{3}\cos\theta_{\pi\pi})\,. 
    \label{eq:diff-Xsect-ee}
\end{equation}
Here $A$ can be expressed in terms of Standard Model (SM) parameters and is explicitly
given in Ref.~\cite{Abel:1992kz}. The factor of $\frac{1}{3}$ is due to the 
averaging over the $Z^0$ spin states. $\theta_{\pi\pi}$ denotes the angle between 
the pion momenta $\hat{\mathbf{p}}_{\pi^\pm}$. We can now define a normalized correlation 
function
\begin{equation}
    P_{\mathrm{QM}}(Z\to\tau\tau)
    \equiv \frac{d\sigma/d\cos\theta_{\pi\pi}(e^+e^-\to\pi^+\bar\nu_\tau\pi^-\nu_\tau)}{\sigma(e^+e^-\to\pi^+\bar\nu_\tau\pi^-\nu_\tau)}=\frac{1}{2}\left(1-\frac{1}{3}\cos\theta_{\pi\pi}\right)\,.
    \label{eq:P_QM}
\end{equation}

Bell's inequality for correlated spin-1/2 particles \cite{Bell:1964kc} is given as
\begin{equation}
\label{eq:bell_orig}
    1+P(\vec b, \vec c)\geq \left|P(\vec a, \vec b) -P(\vec a, \vec c)\right|.
\end{equation}
Here $P(\vec a, \vec b)$ is the correlation function measured with spin-analyzer
settings $\vec a$ (on the left) and $\vec b$ (on the right). Recall that in the LEP 
experiment the pion momenta (in their respective parent $\tau$ rest frames) are the 
purported $\tau$ spin analyzers. If we insert our result for $P_{\mathrm{QM}}(Z\to\tau
\tau)$ from Eq.~\eqref{eq:P_QM} into Bell's inequality we find that it is always satisfied,
\textit{i.e.,} for all angles. However, the reaction $e^+e^-\to Z\to \tau^+\tau^-$ does not 
produce a fully spin-singlet configuration\,\cite{Tsai:1971vv}. Therefore, the correlation
Eq.~\eqref{eq:P_QM} should be checked against the CHSH inequality, given as 
\begin{eqnarray}
    |P(\vec{a},\vec{b}) - P(\vec{a},\vec{b'}) + P(\vec{a'},\vec{b}) + P(\vec{a'},\vec{b'})|\leq 2\,.
    \label{eq:chsh}
\end{eqnarray}
We find that the CHSH inequality is also satisfied for $P_{\rm QM}$ from 
Eq.\,(\ref{eq:P_QM}) for all angles.

We now address the question of \textit{why} the correlation function satisfies the CHSH 
inequality. It turns out that, by adapting the construction of Kasday 
\cite{kasday1971experimental} to the collider context, we can construct an LHVT which 
reproduces the quantum mechanical prediction for the reaction at hand $e^+e^-\to\pi^+\bar
\nu_\tau\pi^-\nu_\tau$. To see this, consider the full differential cross section over 
the solid angles $\Omega_\pm$ corresponding to the unit momentum vectors $\hat{\mathbf{p}}
_{\pi^\pm}$, respectively
\begin{equation}
    \frac{d\sigma}{d\Omega_+d\Omega_-}(e^+e^-\to\pi^+\bar\nu_\tau\pi^-\nu_\tau)
    \equiv f(\hat{\mathbf{p}}_{\pi^+},\hat{\mathbf{p}}_{\pi^-})\,.
\end{equation}
We have equated this with $f(\hat{\mathbf{p}}_{\pi^+},\hat{\mathbf{p}}_{\pi^-})$, a 
function only of the unit pion momenta. If we integrate this over all but the 
relative angle between the pion momenta, $\theta_{\pi\pi}$, we retrieve 
Eq.~\eqref{eq:diff-Xsect-ee}. Now let the hidden variables (in the sense of Bell 
\cite{Bell:1964kc}) for each $\tau$ emitted by the $Z^0$ be a set of unit vectors 
\begin{equation}
    \{\hat{\boldsymbol{\lambda}}_e,\,\hat{\boldsymbol{\lambda}}_\mu,\,\hat{\boldsymbol{\lambda}}_\pi,\,\hat{\boldsymbol{\lambda}}_\rho,\ldots \}\,.
\end{equation}
In the case of the hidden variable theory each $\tau$ emitted by the $Z^0$ carries
with it one such $\hat{\boldsymbol{\lambda}}_i\in\{\hat{\boldsymbol{\lambda}}_e,\,
\hat{\boldsymbol{\lambda}}_\mu,\,\hat{\boldsymbol{\lambda}}_\pi,\,\hat{\boldsymbol
{\lambda}}_\rho,\ldots\}$, depending on which mode it should decay into. For the 
decay $\tau^\pm\to\pi^\pm\nu_\tau$ it is the $\hat{\boldsymbol{\lambda}}_{\pi^\pm}$ 
and this hidden variable tells it to decay such that
\begin{equation}
    \hat{\mathbf{p}}_{\pi^\pm}\stackrel{!}{=}\hat{\boldsymbol{\lambda}}_{\pi^\pm}\,.
    \label{eq:lam=p}
\end{equation}
Let us denote the original probability distribution of the hidden variables according 
to which the $\tau$'s are emitted as
\begin{equation}
F(\hat{\boldsymbol{\lambda}}_{\pi^+},\hat{\boldsymbol{\lambda}}_{\pi^-})\,.
\end{equation}
Next we identify $F$ with $f$,
\begin{equation}
    F(\hat{\boldsymbol{\lambda}}_{\pi^+},\hat{\boldsymbol{\lambda}}_{\pi^-}) 
    \stackrel{!}{=}f(\hat {\mathbf{p}}_{\pi^+},\hat{\mathbf{p}}_{\pi^-})\,,
\end{equation}
assuming the relation Eq.~\eqref{eq:lam=p}. This is just the full differential cross 
section, which we know agrees with the data. Thus the differential cross section as 
computed in the quantum field theory of the SM can be considered an LHVT. This LHVT 
is by construction local and generates separable correlations ({\it cf.}~Appendix~\ref{app:separability})\,\footnote{For a discussion of the 
non-entanglement of LHVTs in the context of the paper by Werner \cite{Werner:1989zz} 
see the Summary and Conclusions in Section~\ref{sec:conclusion}.}~and it necessarily 
satisfies the CHSH inequality. As stressed in Ref.~\cite{Abel:1992kz}, this 
construction is only possible because all momentum components commute,
\begin{eqnarray}
    [(\hat p_{\pi^\pm})_i, (\hat p_{\pi^\pm})_j]&=& 0\,,\;\forall\, i,j \, ,\\[2mm]
    [(\hat p_{\pi^+})_i, (\hat p_{\pi^-})_j]&=& 0\,,\;\forall \, i,j\,.
\end{eqnarray}
\textit{Only in this case} does quantum mechanics provide the required function $f(\hat
{\mathbf{p}}_{\pi^+},\hat{\mathbf{p}}_{\pi^-})$ that we can trivially match to $F(\hat
{\boldsymbol{\lambda}}_{\pi^+},\hat{\boldsymbol{\lambda}}_{\pi^-})$. For non-commuting 
observables this function does \textit{not} exist. For example, in the two-photon 
experiments \textit{e.g.,} by Aspect et al. \cite{Aspect:1982fx}, there is no such 
function for the independent spin components $S_{x,y}$ of the photons. It is then 
not possible to construct the corresponding LHVT via this method.

Note it is possible to extract the $\tau\tau$ spin correlations $P_{\sigma^\tau\sigma
^\tau}=-\cos\theta_{\pi\pi}$ from $P_{\rm QM}(Z\to\tau\tau)$. For this one must perform 
the following mathematical manipulation,
\begin{equation}
P_{\sigma^\tau\sigma^\tau}\equiv6\left[P_{\mathrm{QM}}(Z\to\tau\tau)-\frac{1}
{2}\right] =-\cos\theta_{\pi\pi}\,.
\end{equation}
$P_{\sigma^\tau\sigma^\tau}$ then violates Bell's inequality. However, the 
connection between $P_{\mathrm{QM}}(Z\to\tau\tau)$ and the physical quantity 
$P_{\sigma^\tau\sigma^\tau}$ only holds within quantum field theory or quantum 
mechanics. But quantum mechanics is what one wants to test in this context. 
It is therefore circular logic to extract $P_{\sigma^\tau\sigma^\tau}$ from 
$P_{\mathrm{QM}}(Z\to\tau\tau)$ using quantum mechanics and then to employ it 
to test quantum mechanics. This connection between the measured angular 
correlations and the inferred underlying spin correlations is discussed in more 
detail in Ref.~\cite{Bechtle:2025ugc}.

%% file: tex/04_analysis/analysis.tex


\section{Examples of Fermion Pair Production at Colliders}
\label{sec:analysis}

In this section, we extend the above analysis to several benchmark reactions for 
studying locality via Bell's inequality, as well as entanglement at the LHC. 
For this we perform a 
numerical analysis, as analytic computations of the full hadronic cross sections 
including the parton density functions is not possible. We first briefly revisit 
the process $e^+ e^-\to Z^0\to\tau^+\tau^-\to\pi^+\pi^- \nu_\tau \bar{\nu}_\tau$ 
\cite{Abel:1992kz} from the previous section in order to demonstrate our numerical 
methods. Following that, we  discuss the more recent examples of $\tau^+\tau^-$ 
and $t\bar{t}$ production, as well as Higgs boson decays at the LHC. We generate 
the processes using \texttt{MadGraph5\_aMC@NLO} \cite{Alwall:2014hca} in order to 
preserve the full spin correlation between the produced particles. The spin 
correlation of the produced particles can be inferred from the angular 
distributions of their visible decay products, however only when presupposing 
quantum mechanics, see also Ref.~\cite{Bechtle:2025ugc}.

\subsection{Revisiting $e^+ e^- \to Z^0\to\tau^+ \tau^-\to\pi^+\bar\nu_\tau\pi^-\bar 
\nu_\tau$ at LEP}

Here, we consider the two-body decays $\tau^\pm \to \pi^\pm \nu_\tau$. The correlation 
between the spins of the two $\tau$ leptons can be studied by looking at the 
\begin{figure}[h!]
    \centering
    \includegraphics[width=0.85\textwidth]{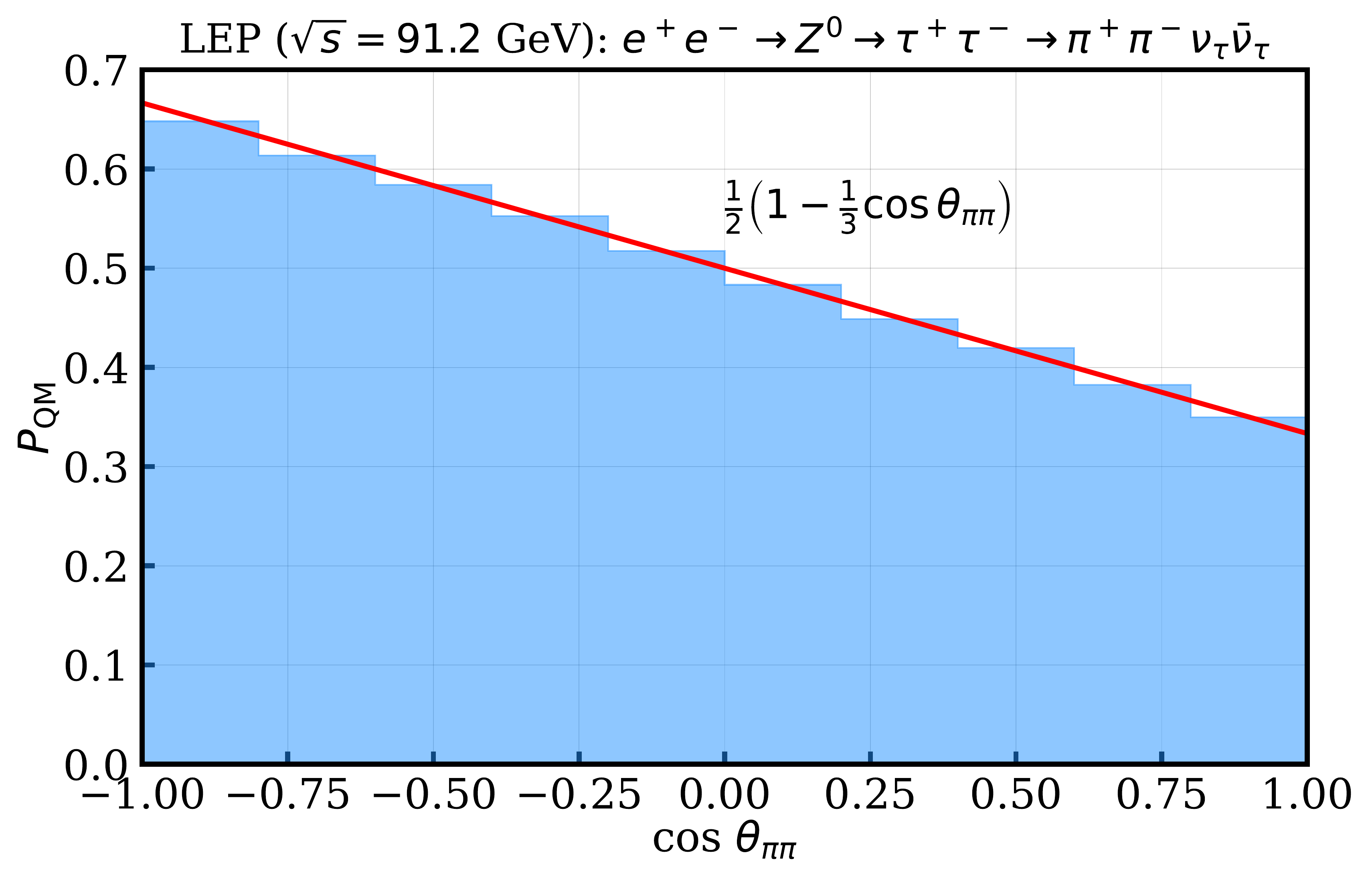}
\caption{A numerical evaluation of the histogram for the correlation function $P_{\mathrm{QM}}
(Z\to\tau\tau)$ as a function of $\cos\theta_{\pi\pi}$, the cosine of 
the angle between the two final state charged pions. The red line denotes the 
analytical prediction \cite{Abel:1992kz}, shown here also as the formula corresponding to Eq.~\eqref{eq:P_QM}.
    }
    \label{fig:LEP_tautau}
\end{figure}
differential cross section as a function of the cosine of the angle between the two 
pion momenta ($\cos\theta_{\pi\pi}$), each boosted to the respective rest frames of 
the corresponding parent $\tau$. In Eq.~\eqref{eq:diff-Xsect-ee} we presented an 
analytic result for the differential cross section \cite{Abel:1992kz}. Using a 
modification of the Kasday construction \cite{kasday1971experimental}, we 
furthermore proved that this differential cross section as a function of the unit pion 
momenta, $\hat{\mathbf{p}}_{\pi^\pm}$, constitutes an LHVT and as such should satisfy 
the CHSH inequality,  once properly normalized. The relevant parameter here is $\cos\,
\theta_{\pi\pi}$, the cosine of the angle between $\hat{\mathbf{p}}_{\pi^-}$ and $\hat 
{\mathbf{p}}_{\pi^+}$. In order to validate our numerical setup, we reproduce this 
result first using the aforementioned \texttt{MadGraph5}. The distribution of the 
normalized correlation function $P_{\mathrm{QM}}(Z^0\to\tau^+\tau^-)$ given in 
Eq.~\eqref{eq:P_QM} as a function of $\cos\,\theta_{\pi\pi}$ is shown in 
Fig.\,\ref{fig:LEP_tautau}. Our numerical study matches with the analytic result of 
Ref.\,\cite{Abel:1992kz} shown here as the {\it red line}. This satisfies the CHSH 
inequality for all angles. This confirms our previous results.

In Refs.~\cite{Han:2025ewp, Ehataht:2023zzt} this process is studied at other 
$e^+e^-$-colliders, the Beijing Electron Positron Collider (BEPC) and Belle II, respectively.
Both of these colliders run at lower energies, where the cross section is dominated by photon 
exchange. This does not affect the arguments. The authors claim to be able to test for 
locality via Bell's inequality depending on the experimental uncertainties. We disagree with 
this assessment for the stated reasons. Furthermore, tests of entanglement are only possible 
within the context of the Standard Model or an alternative quantum field theory.

See also Ref.~\cite{Yang:2026uwu} on a recent study of the same process of tau-lepton 
pair production at lower energies at the Super Tau-Charm Facility, \textit{i.e.} dominantly 
mediated by a photon instead of a $Z^0$. They agree that there is no test of locality via 
Bell's inequality and one can only test for entanglement within the framework of the 
Standard Model.
\subsection{$\tau^+ \tau^-$ from Higgs Boson Decays at the LHC}

Moving from LEP to the LHC, we can also study the correlation of $\tau$ leptons 
coming from Higgs boson decays, $H\to\tau^+\tau^-$. We consider the dominant 
gluon-gluon fusion (ggF) production mode for the Higgs boson, $H$, taking the 
LHC center-of-mass energy to be 14\,TeV. We use the Higgs Effective Field Theory 
(HEFT) model for this production, with the corresponding 
UFO\,\cite{Degrande:2011ua,heft} input in \texttt{MadGraph5}. The full 
differential cross section as a function of the unit pion momenta $\hat{\mathbf
{p}}_{\pi^\pm}$ is denoted by
\begin{equation}
    \frac{d\sigma}{d\Omega_{\pi^+}d\Omega_{\pi^-}}
    (pp\to H\to\tau^+\tau^-\to\pi^+\bar\nu_\tau\pi^-\nu_\tau)
    \equiv f_{\tau\tau}(\hat{\mathbf{p}}_{\pi^+}, \hat{\mathbf{p}}_{\pi^-})\,.
\end{equation}
Here, as before, the $\hat{\mathbf{p}}_{\pi^\pm}$ are determined in the respective 
rest frame of their parent $\tau^\pm$. Applying our construction of the previous 
section in completely analogous fashion
\begin{figure}[h!]
    \centering
    \includegraphics[width=0.8\textwidth]{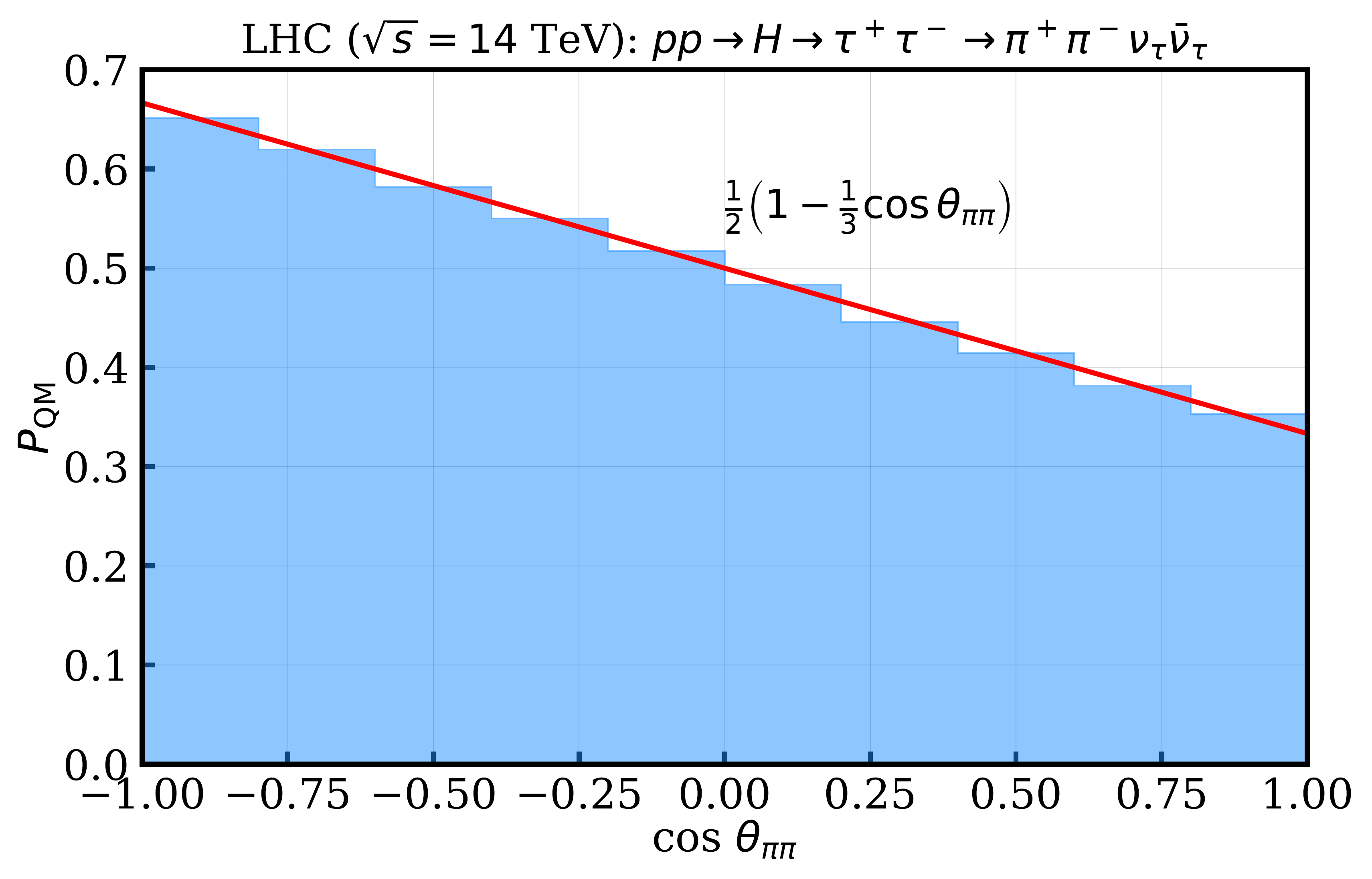}
    \caption{The blue histogram shows the normalized differential cross section for the 
    reaction $pp\to H\to\tau^+\tau^-\to\pi^+\bar\nu_\tau\pi^-\nu_\tau$, at the LHC.
    The red line denotes the numerical fit, which agrees with the previous analytical 
    function, also shown here in the plot.}
    \label{fig:LHC_tautau}
\end{figure}
it is straightforward to show that this function $f_{\tau\tau}(\hat{\mathbf{p}}_{\pi^+}, 
\hat{\mathbf{p}}_{\pi^-})$ constitutes a hlocal idden variable theory, an LHVT. After 
integrating over all but the relative angle between the pion momenta we obtain the 
corresponding normalized correlation function
\begin{equation}
    P_{\mathrm{QM}}(H\to\tau\tau)\equiv \frac{d\sigma/d\cos\theta_{\pi\pi}(
    pp\to H\to\tau^+\tau^-\to\pi^+\bar\nu_\tau\pi^-\nu_\tau)}{\sigma(
    pp\to H\to\tau^+\tau^-\to\pi^+\bar\nu_\tau\pi^-\nu_\tau)}\,.
    \label{eq:P_QM-Higgs-tau-tau}
\end{equation}
It should satisfy Bell's inequality, as the Higgs boson has spin zero and the 
$\tau\tau$-system is thus in a spin singlet state. In Fig.\,\ref{fig:LHC_tautau} we 
show our numerical evaluation of $P_{\mathrm{QM}}(H\to\tau\tau)$ as a function of 
$\cos\theta_{\pi\pi}$. This has exactly the same form as the normalized correlation 
function determined from $\tau$ leptons from $Z$ boson decays at LEP. Indeed
$P_{\mathrm{QM}}(H\to\tau\tau)$ satisfies Bell's inequality for all angles 
$\theta_{\pi\pi}$, as it should since it is an LHVT. 

It should be clear that such an LHVT is local by construction. The whole point of an 
LHVT as defined in Bell's paper \cite{Bell:1964kc} is that the two particles are truly 
separated once they leave the source. Coming from a common source, they of course 
share information. This could for example be linear or angular momentum. However in an 
LHVT \textit{no} information is in any way shared, once the particles have \textit{separated}. 
The measurements on either particle are independent; the correlations are thus separable ({\it cf.}~Appendix~\ref{app:separability}). 
By construction, the function $f_{\tau\tau}(\cos\theta_{\pi\pi})$ describes the data, 
as well as the Standard Model, \textit{cf.} Fig.\,\ref{fig:LHC_tautau}. Since the data 
can \textit{also} be described by a classical, local, and separable theory without entanglement, 
it is not possible to test for locality via Bell's inequality. Similarly, an entanglement test
is only possible within the Standard Model or an alternative quantum field theory.

This signature has been investigated at future $e^+e^-$ colliders in 
Ref.~\cite{Altakach:2022ywa, Ma:2023yvd,  Breuning:2024wxg}, \textit{i.e.} $e^+e^-\to H\to
\tau^+\tau^-$. In Ref.~\cite{Altakach:2022ywa} the authors consider a test of locality via 
Bell's inequality. They acknowledge the work in Ref.~\cite{Abel:1992kz} and its implications 
but incorrectly neglect that the determination of the analyzing power of the decay $\tau\to
\pi+\nu$, though well known, itself assumes quantum field theory. See 
Ref.~\cite{Bechtle:2025ugc} for a more detailed discussion of this point. 
Ref.~\cite{Ma:2023yvd} considers tests of locality via Bell's inequality which we believe do 
not apply for the stated reasons. Ref.~\cite{Breuning:2024wxg} only consider measurements of 
entanglement. These are only valid within the Standard Model. We briefly discuss the related 
paper of Werner \cite{Werner:1989zz} in the Summary and Conclusions, Section~\ref{sec:conclusion}.

\subsection{Spin Correlations in the $t \bar{t}$ System at the LHC}
\label{sec:ttbar}

We next consider the $t\bar t$ system at the LHC. This has recently been widely 
studied as a potential test of locality via Bell's inequality and/or entanglement, 
see for example Refs.~\cite{
Afik:2020onf, Fabbrichesi:2021npl,  
Severi:2021cnj, 
Aoude:2022imd,
Afik:2022kwm, 
Aguilar-Saavedra:2022uye, 
Fabbrichesi:2022ovb,
Afik:2022dgh,
Severi:2022qjy,
Dong:2023xiw,
Aguilar-Saavedra:2023hss,
Han:2023fci,
Cheng:2023qmz, 
Aguilar-Saavedra:2024fig,
Maltoni:2024tul,
Aguilar-Saavedra:2024hwd,
Aguilar-Saavedra:2024vpd,
Maltoni:2024csn,
White:2024nuc,
Dong:2024xsg,
Cheng:2024btk,
Dong:2024xsb,
Han:2024ugl, ATLAS:2023fsd, CMS:2024pts, 
CMS:2024zkc}. 
In these studies, $t\bar{t}$ production and the subsequent di-leptonic decay of the 
top quarks, $t\to b\ell^+\nu_\ell,\,\bar t\to\bar b\ell^-\bar\nu_\ell\,$, is considered. 
They analyze the correlation between the momenta of the two final state charged leptons. 
Note that unlike the previous examples, as well as those that follow below, this is a 
3-body decay of the correlated spin-1/2 particles, here the $t$ and $\bar t$, as opposed to 
a 2-body decay.

Let us start our discussion by introducing two different quantities, which are 
crucial for understanding most of the theoretical analyses, as well as the ATLAS 
and CMS results \cite{ATLAS:2023fsd, CMS:2024pts, CMS:2024zkc}, see for example 
Ref.~\cite{Afik:2020onf}. First is the spin density matrix describing the $t\bar{t}$ 
two-qubit system, which is given by
\begin{equation}
    \rho_{t\bar{t}} = \frac{1}{4}\left(\mathbf{I}_4 + B_i^+\sigma_i \otimes 
    \mathbf{I}_2 + B_i^-\mathbf{I}_2\otimes\sigma_i + C_{ij}\sigma_i\otimes\sigma_j 
    \right)\,,
    \label{eq:density}
\end{equation}
where $\mathbf{I}_n$ is the $n\times n$ identity matrix, $\sigma_i$ are the Pauli 
matrices for the top spins, $\mathbf{B}^+$ and $\mathbf{B}^-$ are the spin 
polarization vectors of the top and anti-top quarks, respectively, and the matrix 
$\mathbf{C}$ encodes the spin-spin correlation between the two particles. The second 
quantity of interest is the normalized differential cross-section of the process,
\begin{equation}
  \!\!\!\!\!\!\!  \frac{1}{\sigma}\frac{d\sigma}{d\Omega_+\Omega_-}(pp\!\to t\bar t\!\to (b\ell^+
    \nu_\ell) (\bar b\ell^-\bar\nu_\ell)) \!=\! 
    \frac{\left(1 + \overline{\mathbf{B}}{}^+\cdot\hat{\mathbf{q}}_+\! - \overline{\mathbf{B}}{}^-\cdot\hat{\mathbf{q}}_- \!- \hat{\mathbf{q}}_+\cdot\overline{\mathbf{C}}\cdot\hat{\mathbf{q}}_- \right)}{4\pi^2}\,,\!\!
    \label{eq:diff_cs}
\end{equation}
where the momentum directions of the final-state antilepton and lepton, in the rest 
frame of their parent top quark, are given by the unit vectors $\hat{\mathbf{q}}_\pm$, 
and $\Omega_\pm$ are the corresponding solid angles, respectively. Eq.\eqref{eq:diff_cs} 
is a parametrization of the normalized differential cross section in terms 
of the measurable quantities $\hat{\mathbf{q}}_\pm$. The quantities $\overline{
\mathbf{B}}{}^\pm$ (3-dimensional vectors) and $\overline{\mathbf{C}}$ ($3\times3$ 
matrix) are just the coefficients of this parametrization. The normalized differential
cross section Eq.~\eqref{eq:diff_cs} is in principle observable at the LHC, and 
from a fit to the measurements one can extract the best values for $\overline 
{\mathbf{B}}{}^\pm$ and $\overline{\mathbf{C}}$. This is a difficult measurement 
\cite{ATLAS:2023fsd, CMS:2024pts, CMS:2024zkc}. 

It is important to note that $\overline{\mathbf{B}}{}^\pm$ and $\overline{\mathbf 
{C}}$ in Eq.~\eqref{eq:diff_cs} are \textit{a priori} distinct from ${\mathbf{B}}{}
^\pm$ and ${\mathbf{C}}$ in Eq.~\eqref{eq:density}. The former are associated with 
the differential cross section, while the latter are parameters of the spin density 
matrix.\footnote{For a more detailed discussion see Ref.~\cite{Bechtle:2025ugc}.}
Unlike in the earlier cited literature, we have thus denoted them by distinct 
symbols. If one assumes quantum field theory, and computes from the spin density 
matrix $\rho_{t\bar{t}}$ in Eq.~\eqref{eq:density} the normalized differential cross 
section $(1/\sigma)d\sigma/d\Omega_+d\Omega_-$ in Eq.~\eqref{eq:diff_cs}, one finds 
that indeed ${\mathbf{B}}{}^\pm$ 
and ${\mathbf{C}}$ can be respectively identified with $\overline{\mathbf{B}}{}
^\pm$ and $\overline{\mathbf{C}}$. However, crucially one has here assumed 
quantum field theory or quantum mechanics in order to make this identification.

\begin{figure}[h!]
    \centering
    \includegraphics[width=0.8\textwidth]{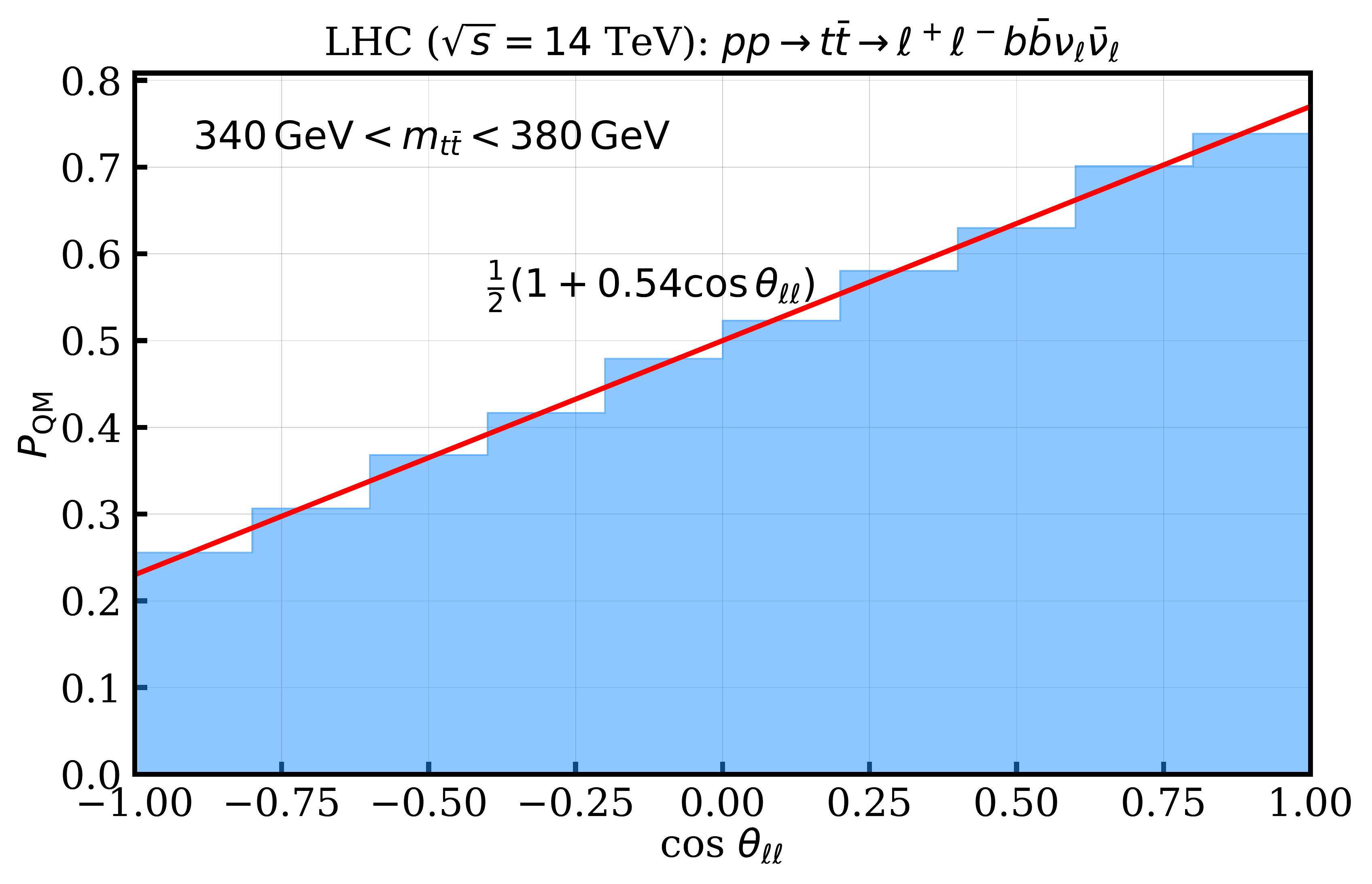}
    \caption{Normalized differential cross section for $t\bar{t}$ production 
    near threshold as a function of $\cos\theta_{\ell\ell}$  at the LHC. The red line denotes the numerical fit.}
    \label{fig:LHC_ttbar}
\end{figure}

Before discussing the case of a possible observation of entanglement as reported by 
ATLAS and CMS, let us first consider the potential test of locality via the CHSH 
inequality in the $t\bar t$ system, as claimed in 
several theoretical papers, see for example Refs.~\cite{Fabbrichesi:2021npl, 
Severi:2021cnj, Aguilar-Saavedra:2022uye, Han:2023fci, Dong:2023xiw, Cheng:2023qmz, 
Cheng:2024btk}. The normalized differential cross section in Eq.~\eqref{eq:diff_cs} 
is a function of the unit lepton momentum vectors, which all commute. Thus we can 
again apply our modified Kasday construction \cite{kasday1971experimental} as in 
Section~\ref{sec:previous_work} to show that the differential cross section is 
itself an LHVT
\begin{equation}
\frac{1}{\sigma}\frac{d\sigma}{d\Omega_+\Omega_-}(pp\to t\bar t\to b\ell^+
    \nu_\ell b\ell^-\bar\nu_\ell)\equiv f_{t\bar t}(\hat{\mathbf{q}}_+,\hat{\mathbf{q}}_-)\,.
    \label{eq:ttbar-ll}
    \end{equation}
That is, the function $f_{t\bar t}(\hat{\mathbf{q}}_+,\hat{\mathbf{q}}_-)$ can be 
considered as a function of the hidden variables $\hat{\boldsymbol{\lambda}}_+=\hat
{\mathbf{q}}_+$, $\hat{\boldsymbol{\lambda}}_-=\hat {\mathbf{q}}_-$, which are 
identified with the decay momenta of the charged leptons, as in Eq.~\eqref{eq:lam=p}. 
Integrating Eq.~\eqref{eq:ttbar-ll} over all but the relative angle $\cos\theta_{\ell
\ell}=\hat {\mathbf{q}}_+\cdot\hat {\mathbf{q}}_-$, we obtain the reduced normalized 
differential cross section 
\begin{equation}
    \frac{1}{\sigma}\frac{d\sigma}{d\cos\theta_{\ell\ell}}(pp\to t\bar t\to b\ell^+
    \nu_\ell \bar b\ell^-\bar\nu_\ell)\equiv \tilde f_{t\bar t}(\cos\theta_{\ell
     \ell}    )\,.
     \label{eq:ttbar-LHVT}
\end{equation}
This LHVT $\tilde f_{t\bar t}(\cos\theta_{\ell\ell})$ should satisfy the CHSH inequality, 
as all LHVTs do. In Fig.\,\ref{fig:LHC_ttbar} we have plotted it as a function of 
$\cos\theta_{\ell\ell}$ as the {\it blue histogram}. In order to compare below with the 
results from ATLAS and CMS, we have here restricted the invariant mass of the $t\bar{t}$ 
pair to be near threshold,
\begin{equation}
340\,\mathrm{GeV}<\,m_{t\bar{t}} \,<380\,\mathrm{GeV}\,.   
\label{eq:threshold}
\end{equation}
It is near threshold that the largest entanglement effects are expected within
quantum mechanics\,\cite{Afik:2020onf}. In Fig.\,\ref{fig:LHC_ttbar} we include a 
linear fit in $\cos\theta_{\ell\ell}$ to the {\it blue histogram} as a {\it red 
solid line}. The coefficients of the fit are shown in the plot. Inserting this 
function, 
we find that it satisfies Bell's inequality for all angles $\theta_{\ell\ell 
}$. This is as expected, as it is an LHVT. Since we have a local function 
which satisfies Bell's inequality for all angles it is \textit{not} possible 
to test for locality via Bell's inequality in this context. This is in 
disagreement with published results in the literature, see for example 
Refs.~\cite{Fabbrichesi:2021npl, Aguilar-Saavedra:2022uye, 
Han:2023fci, Dong:2023xiw, Cheng:2023qmz}. Note, Refs.~\cite{Han:2023fci, 
Dong:2023xiw} consider one top decaying leptonically and the other 
hadronically. The principle of our argument applies analogously.

We find that on inserting the correlation function, $P_{\rm QM}=1/2(1+ 0.54
\cos\theta_{\ell\ell})$, in Eq.\,(\ref{eq:chsh}), the CHSH inequality is also 
satisfied for all angles. In general, if we have a correlation function of the 
form,
\begin{equation}
    P(\vec{a},\vec{b}) = \frac{1}{2}\left(1-c\cos\theta_{ab}\right),
\end{equation}
the CHSH inequality is always satisfied for 
\begin{equation}
-\frac{1}{\sqrt{2}}\leq c\leq \frac{1}{\sqrt{2}},
\end{equation}
since we have $|\cos\theta_{ab}-\cos\theta_{ab'}+\cos\theta_{a'b}+\cos\theta 
_{a'b'}|\leq2\sqrt{2}$. The statements about not testing for locality remain 
the same.

Let us now consider the case for a test of entanglement versus non-entanglement in 
this scenario. Since the normalized differential cross section in 
Eqs.~\eqref{eq:diff_cs}, \eqref{eq:ttbar-ll} is an LHVT, it is inherently local and 
thus generates separable correlations ({\it cf.}~Appendix~\ref{app:separability}).\!\footnote{See also the discussion in 
Section~\ref{sec:conclusion}.} One would thus expect not to have a test of 
entanglement versus non-entanglement here, as in the other examples we have 
discussed. In Ref.~\cite{Afik:2020onf} an entanglement parameter was defined as
\begin{equation}
    D\equiv \mathbf{Tr}[\,\mathbf{C}\,]/3\,,
\end{equation}
based on the matrix $\mathbf{C}$ of Eq.~\eqref{eq:density}. Here $\mathbf{Tr}
[.]$ denotes the trace. A system is 
entangled if \cite{Afik:2020onf}
\begin{equation}
    D< -1/3\,,
    \label{eq:Peres-Horodecki}
\end{equation}
as derived from the Peres–Horodecki criterion \cite{Peres:1996dw, 
Horodecki:1997vt}. The Peres-Horodecki criterion applies specifically to spin 
density matrices, \textit{not} angular correlations of unit momenta. We can 
analogously define, again in our notation,
\begin{equation}
    \overline{D}\equiv \mathbf{Tr}[\,\overline{\mathbf{C}}\,]/3\,,
\end{equation}
based on the matrix $\overline{\mathbf{C}}$ of Eq.~\eqref{eq:diff_cs}. The 
ATLAS collaboration has observed near threshold $t\bar{t}$ production, 
\textit{cf.} Eq.~\eqref{eq:threshold} \cite{ATLAS:2023fsd}. 
They obtain in our notation
\begin{equation}
\overline{D} = -0.537 \pm 0.002 \text{(stat.)} \pm 0.019 \text{(syst.)}\,,
\label{eq:Atlas-exp}
\end{equation}
which is smaller than and significantly differs from $-1/3$. CMS have 
obtained the methodically similar result  $\overline{D} = -0.480^{+0.016}
_{-0.017}\,\text{(stat.)}^{+0.020}_{-0.023}\,\text{(stat.)}$ at the parton level 
\cite{CMS:2024pts}. ATLAS claim their result as a test of entanglement versus 
non-entanglement. We would like to comment on this interpretation. 

Both Refs.~\cite{Afik:2020onf} and \cite{ATLAS:2023fsd} assume that the 
coefficients $\overline{\mathbf{B}}{}^\pm$ and $\overline{\mathbf{C}}$ 
appearing in the differential cross section Eq.~\eqref{eq:diff_cs} are 
respectively identical to $\mathbf{B}^\pm$ and $\mathbf{C}$ appearing in the 
density matrix, \textit{i.e.,}, Eq.\,(\ref{eq:density}). They can then identify
\begin{equation}
    \overline{D}=D\,,
\end{equation}
and thus compare the experimental result Eq.~\eqref{eq:Atlas-exp} to the
entanglement criterion Eq.~\eqref{eq:Peres-Horodecki} derived from the
Peres-Horodecki criterion, leading to an apparent test of entanglement. It
is essential to make this identification, since the Peres-Horodecki criterion,
as we have stated, applies to spin density matrices such as 
Eq.~\eqref{eq:density}, however \textit{not} to differential cross sections 
such as Eq.~\eqref{eq:diff_cs}.

The basis for the assumption of identifying $\overline{\mathbf{B}}{}^\pm$ and 
$\overline{\mathbf{C}}$, respectively with $\mathbf{B}^\pm$ and $\mathbf{C}$
comes from the maximal parity-violating nature of the electroweak charged 
current interactions in the SM, which are responsible for the top quark decay.
Due to this interaction, the final-state leptons carry the spin information of 
their parent top quarks, \textit{i.e.,} the analyzing power is close to 1 
within the SM. See Ref.~\cite{Bechtle:2025ugc} for a more detailed discussion 
of this point. This then allows the reconstruction of the full density matrix 
for the $t\bar{t}$ system by studying the angular distributions of the two 
leptons, so called ``quantum tomography". Once the density matrix is constructed, 
one can check whether the $t\bar{t}$ system is entangled or not. The flaw in 
this argument is that in order to make a connection between the coefficients 
of Eq.\,(\ref{eq:density}) and Eq.\,(\ref{eq:diff_cs}), the underlying theory of 
the SM is necessarily assumed. This is a quantum field theory. If one already 
assumes QM to arrive at the density matrix, one is assuming the existence of 
entanglement for the $t\bar{t}$ system. It is then no surprise that it leads to 
the observation of an entangled state. This is a circular argument. One has to 
assume quantum mechanics to relate the measured angular correlation to the 
desired spin correlation. Thus this is \textit{not} a test of entanglement 
versus non-entanglement. In addition, a by-construction local and separable LHVT fully describes the data; thus, a theory without entanglement is 
consistent with the observations. This is merely a measurement of $D$ within the 
SM.

Next we connect our computation to the ATLAS results. The quantity $\overline 
{D}$ can be experimentally measured from the slope of the normalized 
differential distribution $(1/\sigma)d\sigma/d\cos\theta_{\ell\ell}$, where
$\theta_{\ell\ell}$ is the angle between the two charged leptons from the top 
decays, boosted to the respective parent top quark's rest frame. We study the 
nature of $(1/\sigma)d\sigma/d\cos\theta_{\ell\ell}$ with the same set of 
cuts as described by the ATLAS collaboration. The distribution in $\cos\,
\theta_{\ell\ell}$ from our numerical simulation matches the one produced by 
the experimental collaboration\,\cite{ATLAS:2023fsd}, as shown in 
Fig.\,\ref{fig:LHC_ttbar}. We extract the value of $\overline D$ to be $-0.54$. 
Given the nature of this simulation we do not quote an error. Our result falls 
within 1$\sigma$ of the experimentally measured value Eq.~\eqref{eq:Atlas-exp}. 
This serves as another validation of our analysis setup. However, as mentioned
above, the distribution shown in Fig.\,\ref{fig:LHC_ttbar} satisfies Bell's and 
the CHSH inequalities. This is no surprise, as the data can again be 
described by an LHVT, as constructed above, \textit{cf.} Eq.~\eqref{eq:ttbar-LHVT}.

For clarity we repeat: if we do not assume QM, we cannot identify $\overline{D}
=D$, and thus we can not estimate the quantity $D$ from the data available from 
colliders. Hence we cannot test whether the initial top quark system is entangled.
If we assume QM and identify $\overline{D}=D$ the logic becomes circular. Entangled 
or not, the correlation observed at the LHC between the directions of the two 
leptons coming from the $t\bar{t}$ system can be explained by a theory without 
entanglement, an LHVT, and satisfies the Bell's and CHSH inequalities.

\section{Higgs Boson Decays to Vector Bosons at the LHC}
\label{sec:H-2-VV}
So far we have only discussed the production of correlated spin-1/2 particles 
at colliders: $Z\to\tau\tau;\,H\to\tau\tau;\, pp\to t\bar t$. In his original 
paper Bell considered just such correlations \cite{Bell:1964kc}. Here we 
discuss another set of benchmark examples for entanglement studies at 
colliders coming from the Higgs boson decays to vector bosons: $H\to VV^*$, $V 
\equiv W^\pm, Z$, at the LHC. Note for the resonant production via the 
Standard Model Higgs boson at least one vector boson is off-shell for $V 
\equiv W^\pm, Z$. We shall keep the discussion brief, as the LHVT 
constructions are completely analogous to the previous cases.

\subsection{The Decay $H\to ZZ^*\to\ell^+\ell^-+\ell^{\prime+}\ell^{\prime-}$}
\label{sec:H-2-ZZ}


We begin with the case of $H\to ZZ^*$, where both the $Z$ bosons decay to pairs 
of charged leptons. We assume $\ell=e,\,\mu$, which can easily be reconstructed. 
Repeating our construction from Section~\ref{sec:previous_work} in analogous 
fashion we first consider the full differential cross section 
\begin{equation}
    \frac{d\sigma}{d\Omega_{\ell^+}d\Omega_{\ell'^+}}(pp\to H\to ZZ^*\to
    \ell^+\ell^-+\ell^{'+}\ell^{'-})=f_{ZZ^*}(\hat{\mathbf{p}}_{\ell}, \hat{\mathbf{p}}_{\ell'})\,,
    \label{eq:diff-Xesct-H-ZZ}
\end{equation}
here written as the function $f_{ZZ^*}(\hat{\mathbf{p}}_{\ell},\hat{\mathbf 
{p}}_{\ell'})$ of the unit lepton momenta $\hat{\mathbf{p}} _\ell,\,
\hat{\mathbf{p}} _{\ell'}$. We have focused here on the 2 positively 
charged leptons. Observationally, the charged lepton momenta are determined 
in the parent $Z$ boson rest frame.  For a virtual $Z^0$ boson this is not 
uniquely defined. In our numerical simulation in the appendix, we reconstruct 
the four-momenta and invariant mass of the di-lepton system coming from the 
decay of the virtual $Z^0$ boson at the parton level. We fix the rest frame of 
the virtual $Z^0$ boson as the rest frame of the reconstructed 
di-lepton system. After identifying the momenta with the corresponding hidden 
variables as in Eq.~\eqref{eq:lam=p}, $\hat{\mathbf{p}}_{\ell}\stackrel{!}{=} 
\hat{\boldsymbol{\lambda}}_{\ell}$, we can then identify the LHVT and the 
differential cross section $F(\hat{\boldsymbol{\lambda}}_{\ell+},\hat 
{\boldsymbol{\lambda}}_{\ell'})  \stackrel{!}{=}f_{ZZ^*}(\hat{\mathbf{p}}
_{\ell}, \hat{\mathbf{p}}_{\ell'})\,.$ Thus it is clear that the differential 
cross section in Eq.~\eqref{eq:diff-Xesct-H-ZZ} is an LHVT. 

After integrating over part of the angular space we obtain the reduced 
differential cross section 
\begin{equation}
 \frac{d\sigma}{d\cos\theta_{\ell\ell}}(pp\to H\to ZZ^*\to
 \ell^+\ell^-+\ell^{'+}\ell^{'-})=\tilde f_{ZZ^*}(\cos\theta_{\ell\ell})\,,
        \label{eq:red-diff-Xesct-H-ZZ}
\end{equation}
with $\theta_{\ell\ell}$ the angle between the momenta of two same-sign 
charged leptons. It is not meaningful to insert this into Bell's inequality, 
as Bell's inequality only applies to particles with 2 spin states, such as 
spin-1/2 ($\tau$, $t$) or massless spin-1 (photons) particles. The $Z^0$ 
bosons are massive spin-1 particles and have 3 spin states $\pm1, 0$. In this 
case one possibility is to employ the Collins-Gisin-Linden-Massar-Popescu 
(CGLMP) inequality\,\cite{Collins:2002sun}. However the CGLMP inequality 
refers directly to spin measurements (in our case with results $\pm1,\,0$) and 
not angles of polarimeter settings, $\hat{\mathbf{a}}, \,\hat{\mathbf{b}}$, 
as for Bell. Thus we can not insert Eq.~\eqref{eq:red-diff-Xesct-H-ZZ} 
into Bell's inequality or the CGLMP inequality.

In principle one should derive an extension of Bell’s inequality for massive 
spin-1 particles which can directly be applied to the differential 
cross-section, rather than the spin density matrix. The full unpolarized 
differential cross section $H\to ZZ^*$ should then be checked against this new 
inequality. This is beyond the scope of this paper.

Instead, in Appendix~\ref{app:H-2-ZZ-TT}, we consider the case of the Higgs 
decay $H\to Z_TZ^*_T$, where $Z_T$ denotes a transversely polarized $Z$ 
boson. This only has 2 degrees of freedom and Bell's inequality can again 
be applied. 
We observe that, in this case, Bell’s inequality is always satisfied.

All the same, we can repeat the argument from the previous sections. The 
full differential cross section Eq.~\eqref{eq:diff-Xesct-H-ZZ} is an LHVT. 
As such it is local and generates only separable correlations ({\it cf.}~Appendix~\ref{app:separability}). Furthermore it 
agrees with all the data. It is therefore \textit{not} possible to test for 
locality or for 
entanglement versus non-entanglement using the reaction $pp\to H\to ZZ^*
\to\ell^+\ell^-+\ell^{'+}\ell^{'-}$ at the LHC. This is in disagreement 
with previous work presented for example in 
Refs.~\cite{Aguilar-Saavedra:2022wam, Fabbrichesi:2023cev, 
Aguilar-Saavedra:2024whi, Ashby-Pickering:2022umy, Bernal:2023ruk, 
Bernal:2024xhm, Ruzi:2024cbt, Wu:2024ovc}. Note in 
Refs.~\cite{Ruzi:2024cbt, Wu:2024ovc} the authors consider the decay $H\to 
ZZ^*$ at a muon collider or at the CEPC, respectively. 
Ref.~\cite{Bernal:2024xhm} considers the decay $H\to ZZ^*$ independently of 
the production process. In all cases our arguments apply analogously.

\subsection{The Decay $H\to WW^* \to \ell^+\nu_\ell \ell^{'-}\bar\nu_{\ell'}$}

Next, we consider the example of $H\to WW^*$ at the LHC, where both the $W$ 
bosons decay leptonically to light leptons. Again at least one of the $W$
bosons is virtual. Note the spin correlations for the decay $H\to W^+W^-$, 
followed by the leptonic decays of of the $W$ bosons, $W\to\ell\nu,\,\ell=e,\,
\mu$ have been very useful in the context of the LHC and contributed to the 
initial discovery of the Higgs boson \cite{Dittmar:1996ss, Dittmar:1996sp, 
Dittmar:1997nea, ATLAS:2012yve, CMS:2012qbp}. As for the $Z$ boson, the $W$ 
bosons are spin 1, and we can not apply Bell's inequality. Here we only briefly discuss the construction of the LHVT, which proceeds as in the 
previous cases. We consider the differential cross section for transversely 
polarized $W$ bosons in Appendix~\ref{app:H-2-WW-TT}.

The reaction is
\begin{equation}
    pp\to H\to WW^* \to \ell^+\nu_\ell \ell'^{-}\bar\nu_{\ell'}\,.
\end{equation}
We can consider the corresponding normalized differential cross section 
\begin{equation}
    \frac{1}{\sigma}\frac{d\sigma}{d\Omega_{\ell^+} d\Omega_{\ell'^{-}}}(pp\to H\to WW^*\to \ell^+\nu_\ell \ell'^{-}\bar\nu_{\ell'}) \equiv f_{WW^*}(\hat{\mathbf{p}}_{\ell^+},\hat{\mathbf{p}}_{\ell'^-})\,,
    \label{eq:diff-Xsec-H-WW}
\end{equation}
where $d\Omega_{\ell^+}$ and $d\Omega_{\ell'^-}$ are the differential solid 
angles of the unit momentum vectors $\hat{\mathbf{p}}_{\ell^+}$ and $\hat 
{\mathbf{p}}_{\ell'^-}$. As before, the function $f_{WW^*}(\hat{\mathbf{p}}
_{\ell^+},\hat{\mathbf{p}}_{\ell'^-})$ is an LHVT. Furthermore the data are 
accurately described by the differential cross section. Thus the data can be 
described by a local, separable function ({\it cf.}~Appendix~\ref{app:separability}). Logically, it is thus not possible 
to test for locality or for entanglement versus non-entanglement via the 
reaction $pp\to H\to WW^*\to \ell^+\nu_\ell \ell^{'-}\bar\nu_{\ell'}$ at the
LHC. This is in disagreement with published results in the literature, see 
for example Refs.~\cite{Barr:2021zcp, 
Ashby-Pickering:2022umy, Fabbrichesi:2023cev, Fabbri:2023ncz}.

%% file: tex/05_cuts_and_Bell/cuts_and_Bell.tex
\section{How Momentum Cuts can lead to a Violation of Bell's Inequalities}
\label{sec:momentum-cuts}

\medskip

We have here considered normalized differential cross sections of various 
processes at colliders and have inserted them into Bell's and the related CHSH
inequality. In order to test these inequalities we are obliged to extract 
contributions that are associated with the spins of states, for example by 
taking only the transverse components of massive vector bosons. There is a 
very general issue with this kind of treatment that greatly confuses the 
discussion, and which we now wish to address. This is the fact that in 
experimentally isolating at colliders those parts of an amplitude that apply 
to such inequalities, it is easy to cause a false violation of them, even when 
(as we have argued thus far) we know that there is a perfectly good LHVT 
description of the process. In this section we discuss this issue. We focus on
the Bell inequality as it is much easier to explicitly see where problems 
arise, although in principle this issue applies to all such inequalities. As 
we will see the issue essentially boils down to the fact that a `cut' on data 
implies there is a third possible outcome in the distributions, namely a 
rejection.   

We can begin to appreciate the difficulty if we revisit the reaction $gg\to WW^*$ at the LHC. This has two relevant contributions
 \begin{eqnarray}
     gg&\to& H\to WW^*\to \ell^+\nu_\ell\ell^-\bar\nu_\ell\,,\qquad\mathrm{resonant,}\\
      gg&\to&WW^*\to \ell^+\nu_\ell\ell^-\bar\nu_\ell\,,\;\;\qquad\qquad\mathrm{continuum,}
 \end{eqnarray}
where the latter for example proceeds via a box diagram in leading 
order \cite{Binoth:2005ua, Binoth:2006mf}. In order to construct a test for locality via Bell's inequality (which has two spin states), it is natural to single out the contribution from the transverse $W$ components as done in \ref{app:H-2-WW-TT}. However, that is just a Gedankenexperiment and in reality one does not have ready access to just the transverse components. Suppose that, to circumvent this, we make a ``fair-sampling'' assumption that the longitudinal $W$ component behaves like the transverse ones, and consider the distribution of the entire resonant contribution to the decay. Then we find a surprise: the cross section with {\it that} contribution appears to violate the corresponding Bell's inequality.

How can this be? The main problem is of course that our ``fair sampling'' 
assumption is clearly wrong. The longitudinal components have a very different angular distribution from the transverse ones, thus in effect they are contaminating our Bell test. Then there is the additional issue that one cannot simply disregard the continuum contribution as the contributions add coherently when computing observables such as a normalized differential cross section. Indeed, the 
full normalized differential cross section for the process 
(including the continuum contribution) would satisfy the purported Bell's inequality (if we were to na\"ively continue to identify momenta with spins). 

Thus it is necessary to isolate the relevant ``Bell's inequality contributions''. 
However, because there are two neutrinos in the final state it is not possible to 
reconstruct the helicities of the $W$'s on an event by event basis for this process (unlike for intermediate $Z$'s with 4 charged leptons in the final state). Therefore to do this one would have to resort to cuts on the momenta or invariant mass. 
For example, the dilepton opening angle in the transverse plane, and the invariant mass are both sensitive to the $W$ helicities, so one can use angular and mass distributions to try and isolate the transverse helicities.

However there are two general and crucial objections to such a procedure. The first is the obvious objection that the momentum distributions that one would be using to isolate helicities in this manner are the very same distributions that one would like to use in the Bell's inequality. In other words, one would again be employing  circular logic by using quantum mechanics to construct the Bell test.

The second objection is more subtle and we will spend the rest of this section discussing it. The general problem is that any kind of cut is a form of data rejection, and data rejection (sometimes referred to as a ``no click event'') is effectively a third outcome. 
In this particular case, the introduction of momentum cuts can itself be responsible for an observed violation of Bell's inequality. This is a manifestation of the so-called ``detection loophole" first discussed in Ref.~\cite{Pearle:1970zt}, although in  the present context it appears in an unusually clear and illustrative form that is worth exploring in more depth.

Let us demonstrate the essential issue by following the construction of 
Bell as follows. Recall that our LHVT for the joint probability 
distribution can be written (with hidden variables denoted generically 
by \(\lambda\)) as follows: 
\begin{equation}
    P(\hat {\mathbf p}_a,\hat {\mathbf p}_b)  ~ = ~ \int d\lambda\, F(\lambda)
    \,P(\hat {\mathbf p}_a \mid \lambda) \,P(\hat {\mathbf p}_b \mid \lambda)\,,
\end{equation}
where $\hat{\mathbf{p}}_a$ and 
\(\hat{\mathbf{p}}_b\) are the momentum unit vectors. Note that the integrand factorizes due to locality, with \(F(\lambda)\) being the hidden-variable 
probability density. The distributions of interest here are of the form  
\begin{equation}
P(\hat{\mathbf{p}}_a, \hat{\mathbf{p}}_b) ~=~ \frac{1}{2} \left( 1 - c ~ \hat{\mathbf{p}}_a \cdot \hat{\mathbf{p}}_b\right)\,,
\end{equation}
which encodes correlations between the unit momentum vectors 
$\hat{\mathbf{p}}_a$ and $\hat{\mathbf{p}}_b$ over the sphere, with 
the constant $c$ taking various values, for example  $c=1/3$. The 
normalization is $1/2$ from integrating the cosine over the solid 
sphere, \textit{cf.} Eq.~\eqref{eq:diff-Xsect-ee}.

First let us see how such a distribution can arise from a different kind of 
LHVT from those that we have considered thus far. In the Kasday construction 
\cite{kasday1971experimental}, the hidden variables would trivially correspond 
to the $\hat{\boldsymbol{\lam}}_{a,b}\,$, while we have $P(\hat{\mathbf{p}}
_{a,b}\mid \hat{\boldsymbol{\lam}}_{a,b})= \delta(\hat{\mathbf{p}}_{a,b}- 
\hat{\boldsymbol{\lam}}_{a,b})$ and $F= P(\hat{\boldsymbol{\lam}}_{a},
\hat{\boldsymbol{\lam}}_{b})$. However, consider the alternative LHVT, in 
which the hidden variable consists of just a single unit vector $
\hat{\boldsymbol{\lam}}$ with the integral being over the unit sphere, and with 
the distribution being uniform such that 
\begin{equation}
F(\hat{\boldsymbol{\lambda}}) ~=~ \frac{1}{4\pi}\,.
\end{equation} 
It is a straightforward exercise in integration to show that the response 
function,
\begin{equation}
P(\hat{\mathbf{p}}_a \mid \hat{\boldsymbol{\lambda}}) ~=~ \frac{1}{\sqrt{2}} \left( 1+ i \sqrt{3c} (\hat{\mathbf{p}}_a \cdot \hat{\boldsymbol{\lambda}})\right)\,,
\end{equation}
gives the following joint distribution: 
\begin{align}
P(\hat{\mathbf{p}}_a, \hat{\mathbf{p}}_b) ~&=~ \int d^2 \hat {\boldsymbol{\lambda}} 
\,\frac{1}{8\pi} \left[ 1 -i \sqrt{3c} 
(\hat{\mathbf{p}}_a \cdot \hat{\boldsymbol{\lambda}})
-i \sqrt{3c} 
(\hat{\mathbf{p}}_b \cdot \hat{\boldsymbol{\lambda}})
- 3c 
(\hat{\mathbf{p}}_a \cdot \hat{\boldsymbol{\lambda}})
(\hat{\mathbf{p}}_b \cdot \hat{\boldsymbol{\lambda}})\right] \nonumber \\
\label{eq:distrib}
~&=~\frac{1}{2}  \left( 1 - c~ \hat{\mathbf{p}}_a \cdot \hat{\mathbf{p}}_b \right)\,,
\end{align}
as required. Although the response function is complex, it is allowable for this kind of 
distribution because odd powers of $\hat{\boldsymbol{\lambda}}$ will always vanish in the integral so the probability distribution will be real. 

Bell's original derivation establishes that, for deterministic hidden variables, the 
joint probabilities satisfy inequalities such as the one in Eq.
\eqref{eq:bell_orig}.
In this particular case this translates into  
\begin{equation}
1+ \frac{1}{2}(1-c \cos\theta_{ab}) ~\geq ~{\frac{|c|}{2}} | \cos\theta_{ac} - \cos\theta_{cb} |\,,
\end{equation}
which gives 
\begin{equation}
\frac{3}{|c|} - {\rm sign}(c) \cos\theta_{ab} ~\geq | \cos\theta_{ac} - \cos\theta_{cb} |\,. \end{equation}
As noted in Ref.~\cite{Abel:1992kz}, this is satisfied for all $|c|\leq 1$,
which is always the case since it is required for the joint probabilities to be positive. Indeed, if we suppose 
that the argument in the modulus is positive then we can write this as 
\begin{equation}
    \frac{3}{|c|} ~\geq ~ {\rm sign}(c)\, \hat {\mathbf p}_a \cdot \hat{\mathbf p}_b +  \hat{\mathbf p}_a \cdot \hat{\mathbf p}_c - \hat{\mathbf p}_b \cdot \hat{\mathbf p}_c  \,,
\end{equation}
and the right hand side is clearly less than or equal to 3. (Similarly if the sign of the modulus is negative.)

We can think of the probability in Eq.~\eqref{eq:distrib} as a {\it prior} 
probability, but it is not the probability that one would find after making measurements with cuts on the momenta. Of course we would wish to use the latter to pit the function $P(\hat {\mathbf p}_a,\hat {\mathbf p}_b)$ against the Bell's inequality,  but we will now see that momentum cuts can themselves be responsible for apparent violations of the inequality.
As an example, suppose in an experiment we measure particles along $\hat {\mathbf p}_a$ and in an attempt to make the two particles space-like separated we only accept particles along $\hat {\mathbf p}_b$ that lie in a backwards cone from $\hat {\mathbf p}_a$ with some opening angle $\phi$. 

Then our effective  response function for the second particle is 
\begin{align}    
P(\hat{\mathbf{p}}_b \mid \hat{\boldsymbol{\lambda}}) ~&=~
 \frac{1}{\sqrt{2}} \left( 1+ i \sqrt{3c} \,(\hat{\mathbf{p}}_b \cdot \hat{\boldsymbol{\lambda}})\right) \Theta (\cos\phi - \cos (\pi - \theta_{ab} ) ) \nonumber \\
 ~&=~
 \frac{1}{\sqrt{2}} \left( 1+ i \sqrt{3c} \,(\hat{\mathbf{p}}_b \cdot \hat{\boldsymbol{\lambda}})\right) \Theta (\cos\phi + \cos  \theta_{ab} )\,,
\end{align}
where $\Theta$ is the Heaviside function. Note that there is no asymmetry in the 
integral because we can associate the Heaviside function with $F(\lambda)$ 
instead, which is now conditional on $\theta_{ab}$. The resulting $P(\hat{\mathbf{p}}_a ,\hat{\mathbf{p}}_b ) $ would then be the \textit{posterior} probability in our local hidden-variable theory 
after we have accepted those processes that we think are relevant to space-like 
separated probabilities. It should of course be properly normalised. So we may 
summarize this region-conditioning by instead using the distribution function 
\begin{align}
    F(\hat {\boldsymbol \lambda }) ~=~ \Theta (\cos\phi + \cos  \theta_{ab} )~\frac{1}{2\pi (1-\cos\phi) }~.
\end{align}

One might object to this procedure as it implies that $F$ also knows about the orientation of $\hat{\bf p}_a$ and $\hat{\bf p}_b$ which implies that the data rejection is not local to the detectors at $a$ and $b$, and that this is therefore no longer an LHVT. There are two responses to this. The first is that this is partly our point: momentum cuts are not fair-sampling, and may not respect the locality of different halves of the decay chain. In particular, to reject contributions where the particles are not space-like separated itself requires non-local information. The second response is that, of course one can always devise an LHVT for this new theory by repeating the Kasday procedure. That is, here we would introduce two more hidden variables $\hat{\boldsymbol{\lambda}}_{a}$ and   $\hat{\boldsymbol{\lambda}}_{b}$, and introduce the factors $\delta(\hat{\mathbf{p}}_{a}- \hat{\boldsymbol{\lambda}}_{a})\delta(\hat{\mathbf{p}}_{b}- \hat{\boldsymbol{\lambda}}_{b})  $ 
into the integral (which get associated with the corresponding response functions), and then everywhere that $\cos\theta_{ab}$ appears explicitly in $F$, we would simply replace it by $\hat{\boldsymbol{\lambda}}_{a}\cdot \hat{\boldsymbol{\lambda}}_{b}$. Cuts that depend on only commuting observables can always be implemented into {\it any} LHVT by implementing the Kasday construction in this manner (even if the original LHVT also involves non-commuting observables).  

Continuing then, the joint probability distribution with momentum cuts included becomes 
\begin{equation}
P'(\hat{\mathbf{p}}_a, \hat{\mathbf{p}}_b) ~=~ \frac{\Theta (\cos\phi + \cos  \theta_{ab} )}{1-\cos\phi} \left( 1 - c ~ \hat{\mathbf{p}}_a \cdot \hat{\mathbf{p}}_b\right)\,,
\end{equation}
with the denominator $(1-\cos\phi)$ ensuring that the expression is properly normalized,
\begin{equation}
    \int_{-1}^1  d\cos \theta_{ab} ~P'(\hat{\mathbf{p}}_a, \hat{\mathbf{p}}_b) ~\equiv ~\int_{-1}^{-\cos\phi}   d\cos \theta_{ab} \frac{1}{1-\cos\phi}~ = ~1
\,.
\end{equation}
The denominator is important -- as the acceptance cone is 
made small (\textit{i.e.,} when $\phi\to 0$) the probability density blows up because it is concentrated in the small remaining angle. On the other hand when 
$\phi\to \pi$ the distribution is the original unconstrained one. 

Now the question is whether this distribution should always satisfy Bell's inequality: that is do we expect it to always obey 
\begin{equation}
1 + P'(\hat{\bf p}_a, \hat{\bf p}_b) \stackrel{?}{\geq} | P'(\hat{\bf p}_a, \hat{\bf p}_c) - P'(\hat{\bf p}_c, \hat{\bf p}_b) |\,.
\end{equation}
It is fairly easy to find cases where this inequality is not satisfied. For example take a very small opening angle $\phi$, and let  
$\hat {\mathbf p}_b$ fall {\it outside} this cone (say close to right-angles with the axis along $\hat {\mathbf p}_a$) such that $P'(\hat{\bf p}_a,\hat{\bf p}_b)=0$, but let $\hat {\mathbf p}_c$ be roughly back-to-back with $\hat {\mathbf p}_a$ such that it falls {\it inside} the backwards cone from $\hat {\mathbf p}_a$ but {\it outside} the backwards cone from $\hat {\mathbf p}_b$. Then we have $P'(\hat{\bf p}_a,\hat{\bf p}_b)=P'(\hat{\bf p}_c,\hat{\bf p}_b) =0$, and the Bell's inequality becomes
\begin{equation}
1 ~\stackrel{?}{\geq} ~ \frac{ | 1 - c \cos \theta _{ac}|}{1-\cos\phi }   ~.
\end{equation}
When the angle $\phi \to 0$, the right hand side blows up 
and the inequality is clearly violated for any $c$, and in fact any acute angle $\phi$ (provided the particles fall inside or outside the cones 
as outlined above). 

So in this simple example we see that imposing cuts on the final state phase 
space can produce spurious violations of Bell's inequality in the measurement 
probabilities due to correlations that the cuts themselves introduce into the 
LHVT. As per Ref.~\cite{Pearle:1970zt} on the ``detection loophole", the 
issue is that by rejecting certain outcomes we have effectively added a third 
possible measurement outcome, namely data rejection, but we are erroneously basing our 
probability distribution on only two of them.

It is worth mentioning that the treatment in Ref.~\cite{Pearle:1970zt} was 
based on a more sophisticated configuration than that used above, in the 
sense that it worked by completely local data rejection (or at least data 
rejection that was independent to each observer). This means that the LHVT 
remains an LHVT despite the data rejection. However, this complication was 
actually unnecessary since, as we stressed, one can always construct a new 
LHVT for the theory with momentum cuts, $P'$, by invocation of the Kasday 
construction. 

This conclusion indicates that one can also expect to come across spurious 
violation of Bell's inequality if one only considers or attempts to isolate  
subclasses of the full perturbative calculation of amplitudes of commuting 
variables. For example, requiring a $W$ particle to be off-shell is 
effectively a restriction on the momenta which can introduce correlations. 
Finally, it is worth noting that it is more difficult to devise an experiment 
that induces fake violations of this kind for CHSH type inequalities (the CHSH 
inequality can at best be saturated by na\"ive angular momentum cuts of the 
kind discussed here) but this does not mean they escape this issue.

\section{Summary and Conclusion}
\label{sec:conclusion}

In a previous paper \cite{Abel:1992kz}, we considered the process
\begin{equation}
    e^+e^-\to Z^0 \to \tau^+\tau^- \to (\pi^+\bar\nu_\tau)(\pi^-\nu_\tau)\,.
    \label{eq:ee-tautau-sum}
\end{equation}
Despite the correlation of the $\tau$ spins, there we showed that it can not 
be used as a test of locality via Bell's inequality or the CHSH inequality. 
The arguments we used to prove this were fairly generic. Given that we are 
merely measuring momenta in the final state and momenta all commute, using the 
modified construction of Kasday \cite{kasday1971experimental} it is 
straightforward to show that the normalized differential cross section is 
itself an LHVT. This LHVT necessarily satisfies the CHSH inequality. 
Translating this differential cross section to the underlying $\tau$ spin 
correlation presupposes quantum mechanics, which is not legitimate when
testing quantum mechanics for locality. No statement was made about tests of
entanglement versus non-entanglement.

Here we have now reconsidered the process Eq.~\eqref{eq:ee-tautau-sum} and 
also investigated the reactions
\begin{eqnarray}
    pp&\to&H\to\tau^+\tau^- \to (\pi^+\bar\nu_\tau)+(\pi^-\nu_\tau)\,, \label{eq:proc-H-tautau}\\[2mm]
     pp&\to& t\bar t\to (b\ell^+\nu_\ell) + (\bar b \ell^{\prime-}\bar\nu_{\ell'})\,,\label{eq:proc-gg-tt} \\[2mm]
    pp&\to&H\to ZZ^*\to\ell^+\ell^-+\ell^{\prime+}\ell^{\prime-}\,, \label{eq:proc-H-ZZ}\\[2mm]
    pp&\to&H\to WW^*\to(\ell^+\nu_\ell)+(\ell^{\prime-}\bar\nu_{\ell^{\prime}}) \,,\label{eq:proc-H-WW}
    \label{eq:H-WW-lnulnu-end}
\end{eqnarray}
at the LHC. Using the same line of reasoning as in Ref.~\cite{Abel:1992kz}, we have 
shown that the normalized differential cross section constitutes an LHVT in each 
case. This LHVT is by definition local. It is thus logically not possible to test 
for locality employing these reactions via either Bell's inequality or the CHSH
inequality.

We furthermore explicitly computed the resulting normalized differential distribution 
at the LHC for the processes Eqs.~\eqref{eq:proc-H-tautau}, \eqref{eq:proc-gg-tt} and 
showed that in each case it satisfies Bell's inequality or the CHSH inequality, 
respectively, for all angles between the final state charged leptons or 
pions, respectively. For the case $H\to VV^*$ this is not possible, as 
the $V$ have 3 spin degrees of freedom and Bell's inequality does 
not apply. The CGLMP inequality can also not be employed as it applies directly 
to the spin measurements, not the angular measurements of collider physics. In 
the appendix we have thus considered the pair production of transversely 
polarized vector bosons $H\to V_TV^*_T$. These have 2 spin degrees of freedom 
and one can compare to Bell's inequality. However filtering out the spin 
polarizations effectively corresponds to a cut on the final state phase space. 
In Section~\ref{sec:momentum-cuts} we have shown how this can lead to false 
claims either confirming or violating Bell's inequality. This is a variant of 
the detection loophole for the case of cuts in collider experiments.

\textit{Overall we have found in all cases at LEP or the LHC where it is attempted to 
analyze spin correlations of correlated particles by measuring momenta correlations it is 
\textit{not} possible to test locality via Bell's inequality.} 
We would like to call this a no-go theorem for testing locality via Bell's inequality at 
colliders in accordance with Ref.~\cite{Abel:1992kz}.

In this paper we have gone beyond Ref.~\cite{Abel:1992kz} and have argued that 
the LHVT is by definition local and generates only separable correlations ({\it cf.}~Appendix~\ref{app:separability}). 
This was the original intention in the EPR paper \cite{Einstein:1935rr} and 
also the construction in the paper by Bell \cite{Bell:1964kc}. The emitted 
pair of particles are independent, once they leave the source, \textit{i.e.,} 
their correlations are separable. In each case we have found an LHVT and thus 
a local, separable theory which completely describes the data. It is therefore 
logically not possible to test for entanglement versus non-entanglement in all 
of these scenarios. We have shown in some detail in the $t\bar t$-case how it 
is not possible to connect the measured quantity $\bar D$ as extracted from 
the differential cross section with the entanglement parameter $D$ of the spin 
density matrix. This is only possible by employing quantum mechanics. But it 
is not permissible to use quantum mechanics when testing 
for quantum mechanics. Otherwise one suffers from circular logic. Since 
the arguments we use are again very generic, we go beyond what was 
proven in Ref.~\cite{Abel:1992kz} and also state that: \textit{It is not 
possible to test for locality via Bell's inequality or for entanglement 
versus non-entanglement at colliders, in cases where only final state 
momenta are measured.} We consider this a no-go theorem.

Our discussion is based on the original paper by Kasday 
\cite{kasday1971experimental}, as well as its extension to colliders 
\cite{Abel:1992kz}. The main statement being, that 
an experimental result based on the measurement of commuting variables can 
always be described by an LHVT. This is quite a different statement from the 
seminal result by Werner \cite{Werner:1989zz}. Werner has shown within the 
framework of quantum mechanics that entangled states may exist, so-called 
\textit{Bell-local} states, which however have the property that the 
expectation value of any product of two observables A and B admits a 
description in terms of an LHVT. Thus QM and the LHVT give the same 
experimental predictions on the product of these two observables. Our claim is that, based on the 
collider observations, no distinction can be made between the predictions of a 
separable LHVT and the non-separable entangled QM. This is not in conflict 
with the result by Werner Ref.\,\cite{Werner:1989zz}, in contrast to what is 
pointed out in Ref.\,\cite{Low:2025aqq}.

As an aside we mention: one might think of constructing a purpose built 
experiment, where for example an electron and a positron are produced back to 
back in a singlet state. However it is not possible to measure two independent 
spin components of a free electron, see Refs.~\cite{Mott:1929abc, 
Mott:1965abc} in response to lectures by N. Bohr as quoted on p.699 of 
Ref.~\cite{Wheeler:1984dy}. One could also think of producing $\pi^0$'s on 
resonance and observing the polarization of the emitted photons. However at 
this high energy the polarization is again observed by a scattering event and 
is translated into momenta.

There have also been several proposals to test locality via Bell's inequality using 
entangled pairs of neutral mesons, such as $K^0\bar K^0$ or $B^0\bar B^0$, which can be 
produced at colliders. See for example Refs.~\cite{Lipkin:1968ygs, Lipkin:1988fu, 
DiDomenico:1995ky, CPLEAR:1997jgn, Foadi:1999sg, Pompili:1999tv, Hiesmayr:2000rm, 
Go:2003tx, Caban:2006ij, Belle:2007ocp, Ichikawa:2008eq, Gondran:2013dda, 
Gabrielli:2024kbz}. We have not analyzed these in detail, however we would just point out that in 
Refs.~\cite{Bramon:2004pp, Bramon:2005mg} these have also been refuted via a Kasday
type construction.


%% file: tex/A1_appendix_separability.tex
\section{The Concept of Separability}
\label{app:separability}

In this appendix we expand on the concept of separability, which plays an important role 
throughout the paper.  We first discuss the textbook treatment of separability in the context 
of quantum mechanics, and then extend the definition of a separable correlation to the 
classical case, such as local hidden variable theories (LHVTs), as discussed extensively in 
this paper. This definition is widely used in the literature and we adopt it here.

For the discussion in the context of quantum mechanics we may follow\,\cite{Weinberg_2012}. Let the 
complete set of orthonormal states of a quantum mechanical system $\Psi_m$ be seen by 
one observer $A$ and the complete set of orthonormal states $\Phi_n$ be seen by a 
separate observer $B$. Let the measured quantities only take on discrete values. Furthermore
we assume that $\Psi_m,\,\Phi_n$ are pure states; we shall return to this assumption. Now 
suppose the combined system is in the state
\begin{equation}
\Psi = \sum_{m,n} a_{mn} \Psi_m \otimes \Phi_n\,.
\end{equation}
For example the $\Psi_m$ and $\Phi_n$ could originate from a common source and are 
observed by physically separated observers using detectors. If the coefficients satisfy the 
condition 
\begin{equation}
a_{mn} = b_m c_n\,,
\end{equation}
such that we can write the state $\Psi$ as
\begin{equation}
\Psi = \left(\sum_{m} b_m \Psi_m\right) \otimes \left(\sum_n c_n \Phi_n\right) \equiv \tilde 
\Psi_m \otimes\tilde \Phi_n\,,
\end{equation}
the system is called separable. The expectation value of the product of two operators $A(m),
\,B(n)$ operating separately on the two subsystems is then given by
\begin{equation}
\langle\Psi |A(m) B(n) |\Psi\rangle =\langle  \tilde\Psi_m  | A (m)| \tilde\Psi_m \rangle  \cdot 
\langle \tilde \Phi_n |B(n)|\tilde\Phi_n\rangle \,,
\end{equation}
\textit{i.e.} the expectation value of the two observables factorizes into the individual 
expectation values. Experimentally this means the observations of $A(m)$ are completely 
independent of those for $B(n)$. This is why the overall wave function 
$\Psi$ is called separable. The quantum mechanical system is called entangled if it is 
\textit{not separable}, specifically if the $a_{mn}$ are not simply a product of separate 
functions of only $m$ and $n$, respectively. 

The independence of the observations obtained using physically separated detectors on a 
system of for example two physically separated particles from a common source  is 
considered to always be the case for classical systems, \textit{i.e.} those not governed by 
quantum mechanics. Local hidden variable theories (LHVTs) are classical systems. In the 
original Bell paper \cite{Bell:1964kc} for example, where he gives a 
definition of LHVTs, the expectation value of the joint observation of the two spin-1/2 
particles is given by
\begin{equation}
P(\vec a, \vec b) = \int d\lambda\, F(\lambda)A(\vec a, \lambda) B(\vec b, \lambda)\,.
\end{equation}
Here $A(\vec a, \lambda)$ is independent of the setting $\vec b$ and  \textit{vice 
versa} for $B(\vec b, \lambda)$, while $F(\lambda)$ is simply a probability distribution for the 
hidden variable $\lambda$, with
\begin{equation}
\int d\lambda\, F(\lambda)=1\,.
\end{equation}
The observations $A$ and $B$ being completely independent of each other in this manner  we denote 
 as separable.

In the context of colliders, as here in this paper, the states of interest are 
often impure, and  
described in terms of density matrices, $\hat \rho_A$ and $\hat \rho_B$ for each half of the 
global state. In this formalism the separability of the state is closely aligned with the Bell 
definition of an LHVT. Specifically, a state of the form 
\begin{equation}
\label{eq:density_sep}
\hat \rho ~=~ \sum_\lambda F(\lambda ) ~\hat \rho_A^\lambda \otimes \hat \rho_B^\lambda~,
\end{equation}
where the $\lambda$ sum is over products of different density matrices in the $\hat A$ and 
$\hat B$ Hilbert spaces, respectively, trivially gives the Bell form of the 
expectation value as follows:
\begin{equation}
P(\vec a, \vec b ) ~=~ 
{\rm Tr} (\hat A \, \hat P_{\vec a} \, \hat B \,\hat P_{\vec b} ~  \hat \rho ) 
~=~ \sum_\lambda F(\lambda ) ~  {\rm Tr} (\hat A \,\hat P_{\vec a} ~\rho_A^\lambda)  \,
{\rm Tr} (\hat B\, \hat P_{\vec b}   ~ \rho_B^\lambda) ~,
\label{eq:app-factorizable-density-matrices}
\end{equation}
where $\hat P_{\vec a} $ and $\hat P_{\vec b} $ are projection operators, and where we may 
identify $A(\vec a,\lambda) \equiv {\rm Tr} (\hat A\, \hat P _{\vec a} ~\rho_A
^\lambda) $ and $B(\vec b,\lambda) \equiv \hat {\rm Tr} (\hat B \,\hat P_
{\vec b} ~\rho_B^\lambda ) $ as the local expectation values for a given $\lambda$. Thus 
Eq.~(\ref{eq:density_sep}) is the most general definition of separable quantum theories: all 
states of this form admit an LHVT description, \textit{cf.} 
Eq.~(\ref{eq:app-factorizable-density-matrices}). A state of the form $\hat \rho = \hat \rho_A 
\otimes \hat \rho_B$ is defined as factorizable, but note that for impure states separable does not
necessarily mean factorizable. That is we have 
%
\begin{align}
&\mbox{Pure:} &  \mbox{All states }\supset  \mbox{ Separable }  &\iff & \mbox{Not entangled } &\iff 
&\mbox{Factorizable }     \nonumber \\
&\mbox{Impure:} & \mbox{All states }\supset  \mbox{ Separable }  &\iff & \mbox{Not entangled } &~~~\supseteq &\mbox{Factorizable }     \nonumber 
\end{align}  

As a pertinent example, the density matrix definition of separability leads straightforwardly to the Horodecki-Peres criterion of \cite{Peres:1996dw,Horodecki:1997vt}. Suppose that we have traced over all variables in an experiment except the spins of two particles, such that the two-qubit reduced density matrix takes the form 
\begin{equation}
\rho ~=~ \frac{1}{4} \left(1\otimes 1 + B_i ^+\sigma_i \otimes 1 + B_j^- 1 \otimes \sigma_j+ C_{ij} \sigma_i \otimes \sigma_j  \right)~. 
\end{equation}
The expectation value of spin measurements is given by  $\langle \sigma_k\otimes \sigma_\ell \rangle = {\rm tr} (\sigma_k\otimes \sigma_\ell \,\rho) = C_{k\ell}$. However, if the state is separable as defined by \eqref{eq:density_sep}, this quantity must also be expressible as 
\begin{equation}
C_{k\ell } ~=~ 
{\rm tr } (\sigma_k\otimes \sigma_\ell \,\hat \rho ) ~=~ \sum_\lambda F(\lambda ) ~ {\rm tr }( \sigma_k \hat \rho_A^\lambda ) ~{\rm tr }(\sigma_\ell  \hat \rho_B^\lambda)~.
\end{equation}
The quantities $s^\lambda _{A,k} =  {\rm tr }(\sigma_k \hat \rho_A^\lambda) $ and $s_{B,\ell}^\lambda = {\rm tr }(\sigma_\ell  \hat \rho_B^\lambda)$ 
define two vectors $\vec s^\lambda _{A} $ and $\vec s^\lambda _{B} $
whose modulus is less than or equal to unity. Thus $${\rm Tr} \, C ~=~ \sum_{k}C_{kk} ~=~ \sum _{\lambda} F(\lambda) \vec s_A^\lambda \cdot \vec s_B^\lambda ~\geq ~- \sum _{\lambda} F(\lambda)  |\vec s_A^\lambda || \vec s_B^\lambda| ~\geq~ -1$$ by the Cauchy-Schwarz inequality. We conclude that a separable theory must have \cite{Peres:1996dw,Horodecki:1997vt}
\begin{equation}
{\rm Tr} \, C~\geq ~ -1 ~. 
\end{equation}

In the collider-physics experiments discussed in Sections \ref{sec:analysis} and \ref{sec:H-2-VV}, we do not have detector 
settings. Instead, as an example, we consider the correlations between the final-state charged 
leptons in the $t\bar t$ production process discussed in Section \ref{sec:ttbar}. The normalized distribution 
can be written as
\begin{equation}
\frac{1}{\sigma} \frac{d^6\sigma}{d^3p_+ d^3p_-}\left[pp\to t\bar t\to (b\ell^+\nu_\ell)(\bar b
\ell^{'-}\bar\nu_{\ell^{'-}})\right]\,.
\label{eq:diff-Xsect-separable}
\end{equation}
Here $p_+$ is the 3-momentum of $\ell^+$ and $p_-$ is the 3-momentum of $\ell^{'-}$. This 
describes the correlation of the observations of the momenta $\ell^+$ and $\ell^{'-}$. Note one 
cannot choose a detector setting for each particle at will in these experiments. These 
observations are separable in the following sense. After performing the following integral
\begin{equation}
\int d^3p_-\frac{1}{\sigma} \frac{d^2\sigma}{d\Omega_+ d\Omega_-}\left[pp\to t\bar t\to (b\ell^+\nu_\ell)(\bar b\ell^{'-}\bar\nu_{\ell^{'-}})\right]
\equiv f_+(p_+)
\,,
\end{equation}
one obtains a distribution of events in terms of the final-state $p_+$ momentum 
that is completely independent of the momentum $p_-$. Correspondingly 
alternatively after integrating over $d^3p_+$, one analogously obtains a function 
$f_-(p_-)$, independent of $p_+$. The original double-differential cross section, 
Eq.~(\ref{eq:diff-Xsect-separable}), agrees with the data, and the individual lepton differential 
distributions (obtained by integrating over the respective other lepton variable) 
also agree with the experimental data, implying that the separate integrations are valid. And we 
emphasize again that the distributions $f_\pm(p_\pm)$ are completely independent of each other. 
In this sense we denote the original normalized differential cross section in 
Eq.~(\ref{eq:diff-Xsect-separable}) as separable: the measurements of $p_\pm$ are independent 
of each other. The distribution Eq.~(\ref{eq:diff-Xsect-separable}) constitutes an LHVT and as such 
is a classical correlation of the variables $p_\pm$. We denote these classical correlations as 
separable in the sense defined here.

%% file: tex/04_analysis/analysis_Higgs_new.tex

\section{Higgs Boson Decays to Transversely Polarized Vector Bo\-sons at the LHC}
\label{app:massive_spin1_Bell}

Here we return to the discussion of Section~\ref{sec:H-2-VV}, but restrict 
ourselves to transversely polarized vector bosons. Recall massive vector 
bosons have 3 spin degrees of freedom and Bell's inequality does not apply. 
The Collins-Gisin-Linden-Massar-Popescu (CGLMP) inequality\,
\cite{Collins:2002sun} in principle applies to particles with $n$ spin states. 
However, it is expressed directly in terms of spin measurements, in our case 
resulting in the values $\pm1,\,0$. At a collider we do not measure these spin 
states but only measure angular correlations between the momenta of final 
state particles. That is the crux of the problem we are discussing here, 
\textit{cf.} Ref.~\cite{Bechtle:2025ugc}. Thus the CGLMP inequality also does 
not apply. In contrast transversely polarized massive vector bosons only have
2 spin degrees of freedom and we can again apply Bell's inequality as a test
for locality. However, note the cautionary discussion in 
Section~\ref{sec:momentum-cuts}.

\subsection{The Decay $H\to Z_TZ_T^*\to\ell^+\ell^-+\ell^{\prime+}\ell^{\prime-}$}
\label{app:H-2-ZZ-TT}

We begin with the case of $H\to Z_TZ_T^*$ , where both the $Z$ bosons are transversely polarized and decay to pairs 
of charged leptons. We assume $\ell=e,\,\mu$, which can easily be reconstructed. 
Again, we consider the dominant ggF production of the Higgs boson, using the \texttt{HEFT 
UFO} file containing an effective $ggH$ coupling \cite{Degrande:2011ua,heft}. 
Repeating our construction from Section~\ref{sec:previous_work} in analogous 
fashion it is straightforward to show that the differential cross section 
\begin{equation}
    \frac{d\sigma}{d\cos\theta_{\ell\ell}}(pp\to H\to Z_TZ_T^*\to
    \ell^+\ell^-+\ell'^+\ell'^-)=\tilde f_{ZZ^*}(\cos\theta_{\ell\ell})\,,
    \label{eq:diff-Xsect-H-ZZ-TT}
\end{equation}
is again an LHVT. Here $\theta_{\ell\ell}$ is the angle between the momenta of two 
same sign charged leptons, where the momenta are determined in the rest frame of the
respective parent $Z^0$ boson. As we have mentioned before, for a virtual $Z^0$ boson this is not uniquely
\begin{figure}[hbt!]
    \centering
    \includegraphics[width=0.8\textwidth]{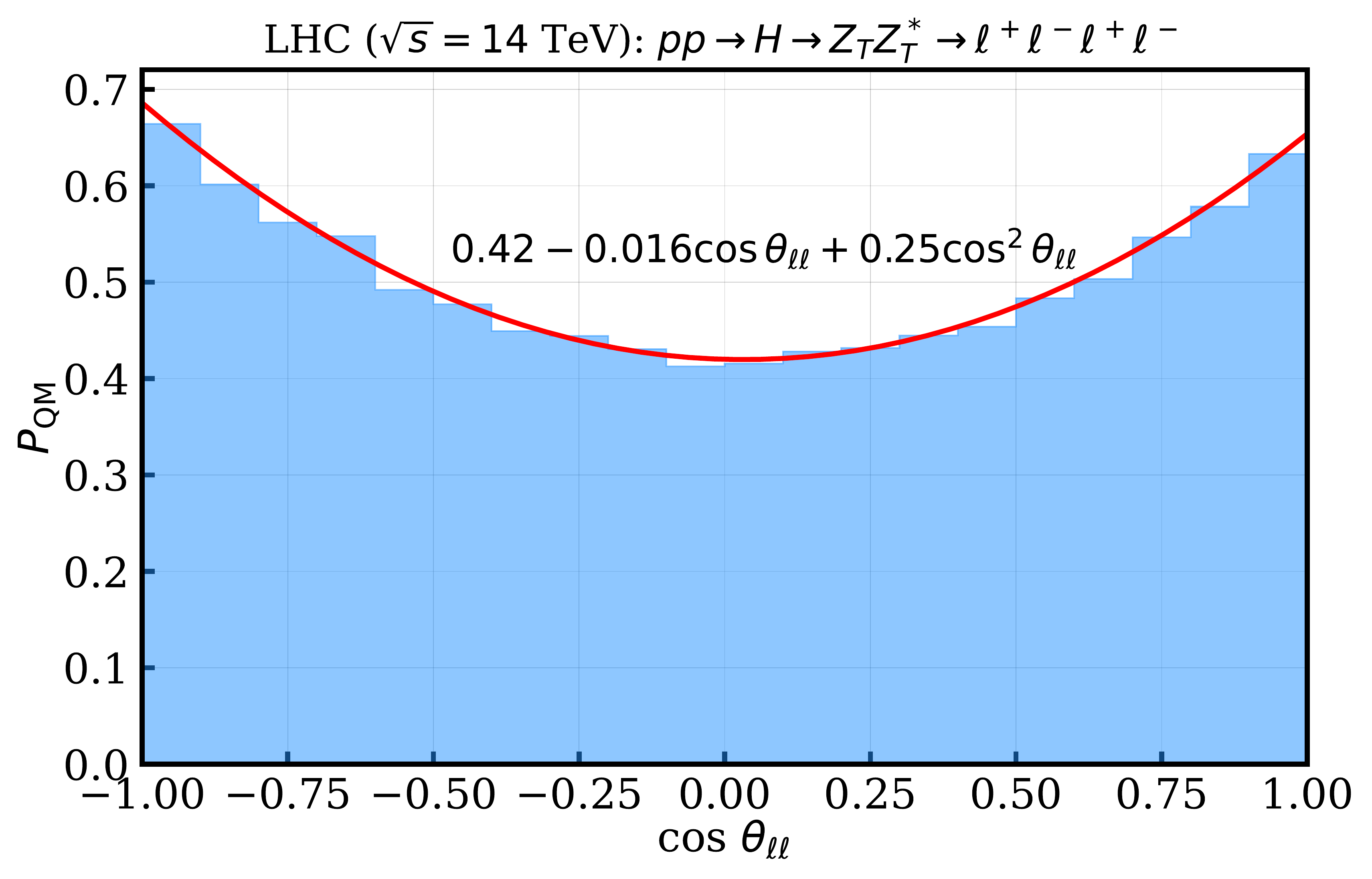}
    \caption{The normalized correlation function $P_{\mathrm{QM}}(H\to Z_TZ_T^*)$ as a function of 
    $\cos\theta_{\ell\ell}$ at the LHC. The red curve is a numerical fit by a second degree polynomial 
in $\cos\theta_{\ell\ell}$. The fit coefficients are displayed in the figure
along the curve.}
    \label{fig:LHC_HtoZZ}
\end{figure}
defined, and we fix the rest frame for the virtual particles following the procedure described in Section\,\ref{sec:H-2-VV}. The normalized correlation 
function is analogously
\begin{equation}
    P_{\mathrm{QM}}(H\to Z_TZ_T^*) \equiv \frac{d\sigma/d\cos\theta_{\ell\ell}(
pp\to H\to Z_TZ_T^*\to\ell^+\ell^-\ell^{'+}\ell^{'-})}{\sigma(pp\to H\to Z_TZ_T^*\to\ell^+
\ell^-\ell^{'+}\ell^{'-})}\,.
    \label{eq:P_QM-Higgs-tau-tau}
\end{equation}
In blue we plot the histogram of this quantity as a function of $\cos\theta_{\ell\ell}$ in 
Fig.\,\ref{fig:LHC_HtoZZ}. The red curve is a numerical fit by a second degree 
polynomial in $\cos\theta_{\ell\ell}$. The fit coefficients are displayed in the 
figure along the curve. We can insert this fitted function into Bell's inequality.
As expected it satisfies Bell's inequality for all angles. Thus it is not 
possible to test locality via Bell's inequality employing this signature.

\subsection{The Decay $H\to W_TW_T^* \to \ell^+\nu_\ell \ell^{'-}\bar\nu_{\ell'}$}
\label{app:H-2-WW-TT}

\begin{figure}[t!]
    \centering
    \includegraphics[width=0.8\textwidth]{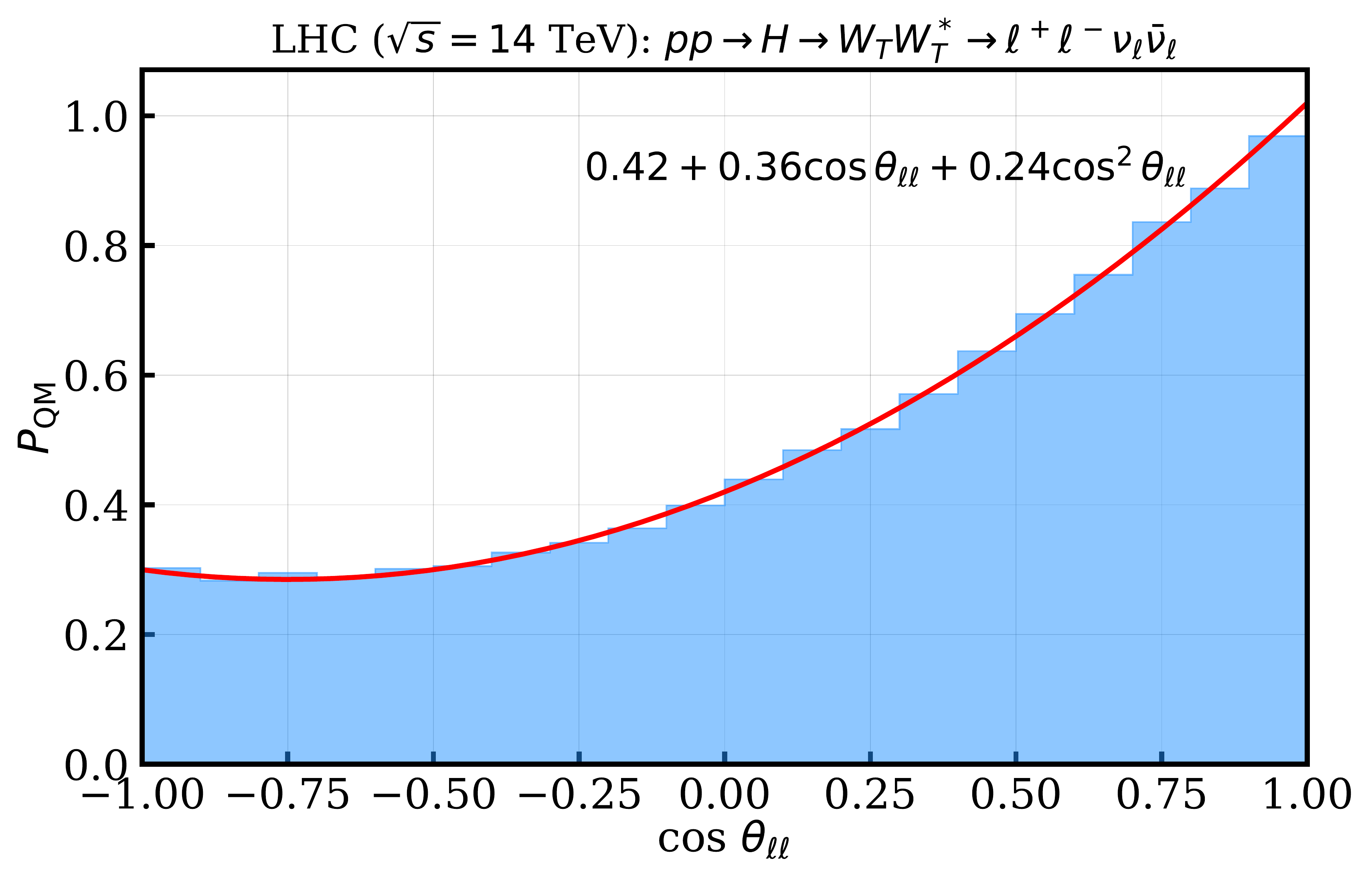}
    \caption{The normalized correlation function $P_{\mathrm{QM}}(H\to W_TW_T^*)$ as a function of 
    $\cos\theta_{\ell\ell}$ at the LHC. The red curve is a numerical fit by a second degree polynomial 
in $\cos\theta_{\ell\ell}$. The fit coefficients are displayed in the figure
along the curve.}
    \label{fig:LHC_HtoWW}
\end{figure}

Next, we consider the example of $H\to W_TW_T^*$ at the LHC, where both the $W$ bosons 
are transversely polarized and decay leptonically to light leptons. Again at least one 
of the $W$ bosons is virtual. For our simulation we again use the \texttt{HEFT UFO} in \texttt{MadGraph}
\cite{Degrande:2011ua,heft}.
The reaction is
\begin{equation}
    pp\to H\to W_TW_T^* \to \ell^+\nu_\ell \ell^{'-}\bar\nu_{\ell'}\,.
\end{equation}
We consider the corresponding differential cross section for this isolated 
process and form the normalized correlation function, as before, \textit{e.g.}
Eq.~\eqref{eq:P_QM},
\begin{equation}
    \frac{1}{\sigma}\frac{d\sigma}{d\cos\theta_{\ell\ell}}(gg\to H\to W_TW_T^*\to \ell^+\nu_\ell \ell^{'-}\bar\nu_{\ell'})\,.
    \label{eq:diff-Xsec-H-WW}
\end{equation}
This is a function of $\cos\,\theta_{\ell\ell}$, where $\theta _{\ell\ell}$ 
is the angle between the two charged final state leptons, determined in 
their respective parent $W$ boson rest frame. The rest frame for the virtual 
$W$ boson is determined in analogous fashion as for the virtual $Z^0$ boson 
in Section~\ref{sec:H-2-VV}. The normalized distribution is shown as a blue 
histogram in Fig.\,\ref{fig:LHC_HtoWW}. The red solid line is the fit via a 
second degree polynomial in $\cos\theta_{\ell\ell}$ to this histogram, where 
the coefficients of the fit are displayed in the graph. Again this function 
satisfies Bell's inequality for all angles, thereby implying that it is not 
possible to test locality via Bell's inequality for this process as well.

\medskip